 \documentclass[preprint]{aastex}

\shorttitle{COS spectroscopy of IC~1613 OB-stars}
\shortauthors{Garcia et al.}

\usepackage{graphicx}
\usepackage[]{natbib}
\usepackage{txfonts}
\newcommand{\Zsun}{$\rm Z_{\odot}$}

\newcommand{\Rsun}{\mbox{$R_{\odot}$}}

\newcommand{\Rstar}{\mbox{$R_{\ast}$}}

\newcommand{\Teff}{\mbox{$T_{\rm eff}$}}
\newcommand{\teff}{\mbox{$T_{\rm eff}$}}
\newcommand{\logg}{\mbox{$\log$~\textsl{g}}~}
\newcommand{\logq}{\mbox{$\log$~\textsl{Q}}~}

\newcommand{\vinf}{\mbox{$v_{\infty}$}}
\newcommand{\vesc}{\mbox{$v_{esc}$}}

\newcommand{\Mdot}{$\dot M$}
\newcommand{\vsini}{$v \sin i$}
\newcommand{\vrad}{$v_{rad}$}

\newcommand{\kms}{~$\rm km \, s^{-1}$}

\newcommand{\halpha}{$\rm H_{\alpha}$}

\begin{document}


\title{Winds of low-metallicity OB-type stars: 
HST-COS spectroscopy in IC~1613.
       }


\author{Miriam Garcia}
\affil{Centro de Astrobiolog\'{\i}a, CSIC-INTA.
Ctra. Torrej\'on a Ajalvir km.4, E-28850 Torrej\'on de Ardoz (Madrid), Spain}
\affil{Instituto de Astrof\'{i}sica de Canarias. 
V\'{i}a L\'{a}ctea s/n, E-38200 La Laguna (S.C. Tenerife), Spain}

\author{Artemio Herrero}
\affil{Instituto de Astrof\'{i}sica de Canarias. 
V\'{i}a L\'{a}ctea s/n, E-38200 La Laguna (S.C. Tenerife), Spain}
\affil{Departamento de Astrof\'{i}sica, Universidad de La Laguna. 
Avda. Astrof\'{i}sico Francisco S\'anchez s/n, E-38071 La Laguna (S.C. Tenerife), Spain}

\author{Francisco Najarro}
\affil{Centro de Astrobiolog\'{\i}a, CSIC-INTA.
Ctra. Torrej\'on a Ajalvir km.4, E-28850 Torrej\'on de Ardoz (Madrid), Spain}

\author{Daniel J. Lennon}
\affil{European Space Astronomy Centre. 
Camino bajo del Castillo, E-28692 Villanueva de la Ca\~nada (Madrid), Spain }

\and

\author{Miguel Alejandro Urbaneja}
\affil{Institute for Astro- and Particle Physics, University of Innsbruck. 
Technikerstr. 25/8, A-6020 Innsbruck, Austria}
\affil{Institute for Astronomy, University of Hawai'i. 
2680 Woodlawn Drive, 96822 Honolulu, USA}





\begin{abstract}
We present the first 
quantitative UV spectroscopic analysis of resolved OB stars in IC~1613. 
Because of its alleged very low metallicity ($\lesssim$1/10 \Zsun, from \ion{H}{2} regions),
studies in this Local Group dwarf galaxy could become a significant step 
forward from the SMC towards the extremely metal-poor massive stars of the early Universe.
We present HST-COS data covering the $\sim$1150-1800\AA~
wavelength range with resolution R$\sim$2500.
We find that the targets do exhibit wind features, and these are similar in strength to SMC  stars.
Wind terminal velocities were derived from the observed P~Cygni profiles
with the SEI method.
The \vinf-Z relationship has been revisited.
The terminal velocity of IC~1613 O-stars
is clearly lower than Milky Way counterparts, but
there is no clear difference between IC~1613 and SMC or LMC analogue stars.
We find no clear segregation with host galaxy
in the terminal velocities of B-supergiants,
nor in the \vinf/\vesc~ ratio of the whole OB star sample
in any of the studied galaxies.
Finally, we present first evidence that the Fe-abundance of IC~1613 
OB stars is similar to the SMC,
in agreement with previous results on red supergiants.
With the confirmed $\sim$1/10 solar oxygen abundances of B-supergiants, our results indicate
that IC~1613's $\alpha$/Fe ratio is sub-solar.
\end{abstract}





\section{Introduction}
\label{s:intro}

The great observatories of this and the coming decade (ALMA, JWST, and E-ELT) 
will bring direct observations of the epoch of re-ionization. 
In the infant Cosmos the first, very massive stars played a crucial role as ionizing sources \citep{Ral10}
and likely progenitors to long-GRBs \citep{Gal09,WH06}. 
An accurate theoretical framework for the evolution of metal-free massive stars, and subsequent calculations 
of stellar yields and ionizing power, are key to interpret the coming observations.
Radiation-driven winds are one of the main pillars of said theory, as agents of mass and momentum removal during the evolution of massive stars.

So far the models for the high-redshift, metal-poor Universe rely on theoretical predictions, 
or on stellar libraries of the Small Magellanic Cloud (SMC).
However, recent results 
indicate that the winds of sub-SMC metallicity massive stars may differ significantly 
from their higher metallicity counterparts. If this is confirmed, massive star evolution and feedback 
would need deep revision, which would then need to be propagated to early Universe models.

In the classical radiation-driven wind framework 
the wind mass loss rate (\Mdot) decreases with decreasing metallicity \citep{VKL00,VKL01,PSL00},
a prediction confirmed by spectroscopic studies of OB stars
down to  SMC metallicities \citep[e.g.][]{Mal07b}.
The scaling relation is $\rm \dot M \propto Z^{\alpha}$,
with $ \rm \alpha \sim 0.7-0.8$ \citep{VKL01,Mal07b}.
The dependence of the wind terminal velocity on metallicity
has only been established empirically and 
is allegedly much weaker, $\rm v_{\infty} \propto Z^{0.13}$~ \citep{LRD92}.
This relation has not been reviewed recently, yet
it was used by \citet{VKL01} and \citet{Mal07b} to derive the \Mdot-Z dependence
(see below).
At extremely low metallicities ($\rm Z \leq 0.01 Z_{\odot}$),
\Mdot~ and \vinf~ are expected to 
decrease with decreasing metallicity more steeply \citep{K02}.

By contrast, very metal poor Luminous Blue Variable stars
with strong optical P~Cygni profiles have been found in the Local Group 
\citep{Hal10} and in farther galaxies \citep{DCS01,pustilnik08,izotov11}.
\citet{Tal11} report six O-type stars with stronger wind momentum than expected
in Local Group galaxies with metallicity $\rm \sim 1/7-1/10 \, Z_{\odot}$.  
We studied an Of star in IC\,1613 possibly exhibiting
a strong wind or, alternatively, anomalous wind acceleration \citep{Hal12}.
These examples challenge the currently accepted metallicity scaling relations for
\Mdot, used by stellar evolution models \citep{MM00,HL00} and ultimately
galactic chemical evolution models (see \citealt{HR10}).
While the analysis with more complete models may explain these findings
(see discussion by \citealt{L12}),
they may constitute the first hints for a more complex wind-driving mechanism.
Better observations on extended samples are necessary to establish or refute this apparent
contradiction to the theory and, in particular,
ultraviolet (UV) spectroscopy covering at least the $\sim$1150-1800\AA~ range is key.

The installation of the Cosmic Origins Spectrograph on-board the Hubble Space Telescope (HST-COS)
has enabled UV spectroscopy of sources a factor of $\sim$ 10 fainter than before.
Access has been enabled to metal-poorer galaxies than the SMC,
which are inconveniently located farther in the Local Group,
but whose stellar population can still be resolved with the HST  \citep{NUVA}. 
Prior to COS, their winds could
only be studied from optical spectral lines (H$_{\alpha}$~ and \ion{He}{2}$\lambda$4686), 
invariant to the wind velocity law 
(parameterized by its exponent $\beta$)
and the terminal velocity unless the wind is very strong.
In fact, the synthetic spectra for these transitions exhibit a degeneracy to a
combination of parameters involving \vinf, \Mdot, and the stellar radius \Rstar~ 
($ Q = \dot M / (v_{\infty} \cdot R_{\ast})^{1.5}$, \citet[][eq. 2]{KP00}).
In addition, the uncertainties of $\beta$~ translate into significant uncertainties in $Q$ (hence \Mdot).
Both \Mdot~ and \vinf~ are required to calculate the modified wind-momentum
($D_{mom} = \dot M \cdot v_{\infty} \cdot R_{\ast}^{0.5}$, 
or equivalently $D_{mom} =  Q \cdot v_{\infty}^{2.5} \cdot R_{\ast}^{2}$).
The correlation between the stellar wind-momentum and luminosity
\citep[the WLR,][]{Kal95} is
our most powerful tool to evaluate radiation-driven wind strength.

Lacking diagnostics for the terminal velocity, optical studies
set \vinf~ from empirical relations to the escape velocity which exhibit large scatter
\citep[e.g. \vinf/\vesc=2.65 for \Teff$\geq$21000~K,][]{KP00}
and then scale it with metallicity using Leitherer's relation. 
This translates into large uncertainties in the actual mass loss rate of the stars, calculated from $Q$,
and the stellar wind momentum.
In particular, the errors of \vinf~ propagate to $D_{mom}$~ so that
$\Delta (\log D_{mom}) = 2.5 \Delta (\log v_{\infty})$, for $Q$ and \Rstar~ constant.

We present first results of our program to study the winds of OB-stars in 
the metal-poor galaxy IC~1613 with HST-COS UV spectroscopy.
Our team is thoroughly characterizing the population of
blue massive stars in this Local Group galaxy with very low metal content 
\citep[0.13 \Zsun~ from \ion{H}{2} regions,][]{Fal07}.
IC~1613 is one of the closest dIrr to the Milky Way
\citep[DM=24.27,][]{Dal01}, with low foreground extinction \citep[E(B-V)=0.02,][]{LFM93}.
Thanks to COS enhanced sensitivity, we can perform ultraviolet spectroscopy
of the targets, and obtain unique insight into their winds.
The wind stratification will be studied from the P~Cygni profiles of metallic lines, 
providing information on \Mdot, $\beta$~ and \vinf.
The degeneracy of optical synthetic spectra to the $Q$ parameter
will be thus broken, and more accurate values will be input into the WLR to assess the winds.



The paper is organized as follows. The HST-COS observations are described
in Section~\ref{s:obs}, and concerns on data reduction in Section~\ref{s:red}.
The morphology of the observed spectra is discussed
in Section~\ref{s:mor} and compared to LMC and SMC counterparts.
Section~\ref{s:sei} details how the terminal velocities were derived,
with results presented in Section~\ref{s:r}.
They allow us to evaluate how \vinf~ changes in Local Group 
galaxies in Section~\ref{s:vinfZ}.
We reflect upon IC~1613 metal content in Section~\ref{s:metal},
and provide our final conclusions in Section~\ref{s:fin}.

\section{Observations}
\label{s:obs}

The list of observed stars is shown in Table~\ref{T:sample},
with identification numbers and photometry from \citet[][hereafter GHV09]{GHV09}.
The sample consists of 8 blue massive stars in IC~1613
selected from previous optical observations 
of our own \citep[][in prep.]{Gal14} or \citet{Fal07}'s catalog.
They were chosen to cover the O- and early-B spectral sub-types,
with representation of the dwarf/giant and supergiant luminosity classes.
The targets have magnitudes
ranging from V $\sim$ 17.5 to 19.5.
They all are isolated to ground-based spatial resolution
within a 2.5\arcsec~ diameter circle (equivalent to COS aperture).


Our program (\dataset[ADS/Sa.HST#12587]{ID: 12587; PI: M. Garcia}) was granted 
23 HST orbits to obtain the COS spectra of the sample stars.
A summary of the observations is provided in Table~\ref{T:log}.
We used the FUV channel for the observations, with the primary science aperture
and the G140L grating centered at $\lambda_c$= 1105\AA, covering $\sim$1118-2250 \AA. 
The resulting spectra have resolution $\Delta\lambda$=0.48\AA~
\citep{COSihb}    
or  $\rm R \sim 2600$ at 1550\AA.
The equivalent $\sim$115\kms~ resolution
in the velocity space suffices
to derive terminal velocities under $\sim$1000\kms~
as shown by \citet{Ual02}.

The used configuration ensured complete coverage of the \ion{N}{5}~$\lambda\lambda$1238.8,1242.8 doublet
which, together with \ion{C}{4}$\lambda\lambda$1548.2,1550.8,
are the only wind signatures seen in the FUV spectra of 
SMC early-O dwarfs \citep{Wal00} and thus the only expected features in their IC~1613 counterparts.
The spectra also cover the additional diagnostic lines 
for Of stars and OB supergiants:
\ion{C}{3}$\lambda$1176, \ion{Si}{4}$\lambda\lambda$1393.8,1402.8,
\ion{He}{2}$\lambda$1640.0 and \ion{N}{4}$\lambda\lambda$1718.0,1718.5.

The observations consisted on 1 visit per target (except \#65426, visited twice)
with several orbits per visit to accumulate exposure time,
aiming at a signal to noise ratio of S/N=20 per resolution element at 1550\AA.
The targets were acquired with the NUV ACQ/IMAGE mode except for
the brightest stars (V$<18.5$),
for which we used the faster ACQ/PEAKXD+PEAKD protocol.
The spectra were taken in TAGFLASH mode, cycling through all
the FP-POS positions for maximum spectral coverage.

\begin{deluxetable}{r l l r r r r r r}
\tabletypesize{\small}
\rotate
\tablecaption{Coordinates and photometric data for the sample stars,
from GHV09. 
Identificators from \citet{Fal07} are provided in column Bal07 for
future cross-reference.
The spectral types provided in column SpT are from \citet{Gal14} except for
\#62024 \citep{Hal12},
\#69217 and \#60449 \citep{Fal07}.
Column Q provides the color index Q=(U-B)-0.72·(B-V).
The color excess E(B-V) was calculated using \citet{winter}'s calibration of intrinsic color 
with spectral type for Galactic stars. \label{T:sample} }
\tablewidth{0pt}
\tablehead{
\colhead{ID}    & \colhead{ID}    & \colhead{SpT} & \colhead{RA[deg]} & \colhead{DEC[deg]} & 
\colhead{V} & \colhead{B-V} & \colhead{Q} & \colhead{E(B-V)} \\
\colhead{GHV09} & \colhead{Bal07} & \colhead{   } & \colhead{J2000.0} & \colhead{J2000.0} & 
\colhead{ } & \colhead{   } & \colhead{ } & \colhead{      } 
}
\startdata                                                                                                              
 69217 & A13     &    O3-O4V((f)) & 16.276065 & 2.178654 & 18.959 & -0.225 & -0.917 & 0.105 \\  
 62024 & \nodata &       O6.5IIIf & 16.252693 & 2.147016 & 19.600 & -0.161 & -0.645 & 0.149 \\
 65426 & B2      & O7.5III-V((f)) & 16.262785 & 2.167927 & 19.621 & -0.202 & -0.906 & 0.118 \\
 67559 & \nodata &   O8.5III((f)) & 16.269862 & 2.156441 & 19.243 & -0.201 & -0.900 & 0.109 \\
 63932 & B7      &           O9II & 16.258220 & 2.134749 & 18.965 & -0.206 & -0.885 & 0.074 \\
 69336 & B3      &           B0Ia & 16.276542 & 2.158706 & 17.686 & -0.152 & -0.870 & 0.088 \\
 62390 & A10     &         B0.5Ia & 16.253812 & 2.178189 & 17.423 & -0.135 & -0.884 & 0.135 \\
 60449 & B4      &         B1.5Ia & 16.248132 & 2.154477 & 18.232 & -0.128 & -0.879 & 0.052 \\  
\enddata
\end{deluxetable}


\begin{deluxetable}{r r r r r r l l r}
\tabletypesize{\small}
\rotate
\tablecaption{Observation summary for the sample stars, including total observing time on target (ToT). \label{T:log}}
\tablewidth{0pt}
\tablehead{
\colhead{ID}    & \colhead{ID}  & \colhead{ROOTNAME} & \colhead{DATE-OBS} & \colhead{SpT} &
\colhead{V} & \colhead{N\_orbits} & \colhead{Acquisition} &
\colhead{ToT} \\
\colhead{GHV09} & \colhead{HST} & \colhead{        } & \colhead{        } & \colhead{   } &
\colhead{ } & \colhead{         } & \colhead{           } &
\colhead{[s]} 
}
\startdata                                                                                                              
 69217 & IC1613-010506-021043 & lbq401070\_x1dsum &2012-07-27&      O3-O4V((f)) & 18.959 & 3 & COS/NUV ACQ/IMAGE        & 6684 \\  
 65426 & IC1613-010503-021004 & lbq402040\_x1dsum &2012-11-03&   O7.5III-V((f)) & 19.621 & 2 & COS/NUV ACQ/IMAGE        & 4284 \\  
       &                      & lbq403040\_x1dsum &2012-11-10&                  &        & 2 &                          & 4284 \\  
 62024 & IC1613-010501-020849 & lbq4040d0\_x1dsum &2012-11-17&           O6.5IIIf & 19.600 & 5 & COS/NUV ACQ/IMAGE        & 11485\\  
 67559 & IC1613-010505-020923 & lbq405070\_x1dsum &2012-11-12&     O8.5III((f)) & 19.243 & 3 & COS/NUV ACQ/IMAGE        & 6684 \\  
 63932 & IC1613-010502-020805 & lbq406040\_x1dsum &2012-11-11&             O9II & 18.965 & 2 & COS/NUV ACQ/IMAGE        & 4284 \\  
 62390 & IC1613-010501-021041 & lbq407010\_x1dsum &2012-07-19&           B0.5Ia & 17.423 & 2 & COS/FUV ACQ/PEAKXD+PEAKD & 4495 \\  
 69336 & IC1613-010506-020931 & lbq408010\_x1dsum &2012-07-19&             B0Ia & 17.686 & 2 & COS/FUV ACQ/PEAKXD+PEAKD & 4495 \\  
 60449 & IC1613-010460-020916 & lbq409040\_x1dsum &2012-07-14&           B1.5Ia & 18.232 & 2 & COS/FUV ACQ/PEAKXD+PEAKD & 4470 \\  
\enddata
\end{deluxetable}




\section{Data reduction}
\label{s:red}

The final calibrated data products for the targets were downloaded from the 
Hubble Data Archive on July 2013,
after ``on-the-fly'' processing by OPUS 2013\_1a and CALCOS  2.19.7.dev23191.
CALCOS calculates the wavelength calibration from the in-exposure lamp flashes,
subtracts the background counts from a predefined detector section,
and computes the target counts.
All the associated exposures within the same visit are then 
combined to produce the final 1-dimensional, flux and wavelength calibrated spectrum.

The software used the latest (as of 2013 May 24) gain-sag reference file,
with improved algorithms for gain-sag removal from the final calibrated spectra. 
Targets \#62390, \#69336 and \#60449, 
observed before the 2nd FUV lifetime position on 2012 July 23,
benefit from this update.

No flat-field correction is applied to the observations in the traditional sense,
since no good quality flat-field image exists with the same configuration  \citep{COSdhb}.
The wavelength dependence of the detector response is instead removed at the flux-calibration step.
The pixel-to-pixel sensitivity variations and the fixed patterns
are corrected to a great extent by using different FP-POS positions 
to construct the final spectrum.
Yet, some large amplitude (up to 10\%) fixed pattern features may remain \citep{COSdhb}.

The accuracy of the spectral absolute flux is diminished by additional factors.
The detector sensitivity decays with time, but the degradation cannot
be well reproduced by the calibration files because it is not linear.
The flux calibration function is poorly constrained at $\lambda <$ 1200\AA~
because of the decreased detector sensitivity, which
translates in uncertainties of 5-10\% for G140L spectra at these wavelengths.
Considering also the uncorrected dip of $\sim$10\% intensity at $\sim$1180\AA~ (but variable location),
the diagnostic lines at the shortest wavelengths of our observations must be used with caution,
especially \ion{C}{3}$\lambda$1176.

Most importantly, the CALCOS extraction procedure
is oblivious to the actual instrumental profile in the cross-dispersion direction.
In other words, the extracted spectra may include some background pixels
or counts from a nearby target.

Regarding the wavelength calibration accuracy,
a common limitation to all COS observing modes is the broad-winged point-spread function.
At the low resolution of our G140L spectra, however, this effect is expected to be small \citep[][Section 5.4]{COSihb}.
More specifically, there may be shifts in the dispersion direction
due to drifts right after moving the 
G140L grism 
and to the accuracy in centering the target in the science aperture along the dispersion direction
\citep{COSdhb}.
We note that we did not detect such drifts between different consecutive exposures (see Sect.~\ref{ss:var}).

Finally, small local deviations from CALCOS's wavelength calibration relations have been reported.
They are suspected to be caused by localized inaccuracies in the geometric correction \citep{COSdhb}.
The departures have not been quantified yet, 
but the 150\kms~ wavelength accuracy requirement for G140L spectra (equivalent to $\Delta \lambda$=0.75\AA~ at 1500\AA), 
is routinely met according to the reports \citep{COSihb}.

We have encountered radial velocity inconsistencies in our data
likely caused by the reported wavelength calibration issues.
The maximum discrepancy within the same spectrum is $\sim$100\kms,
well enclosed in the wavelength accuracy specifications.
More details are provided in Sect.~\ref{ss:com}.

\subsection{Variability and multiplicity}
\label{ss:var}

Taking advantage of HST-COS spatial resolution (higher than the available
ground-based data), we examined the observations
looking for multiplicity.
We first looked for additional targets within the science aperture.
We examined the acquisition files of
the stars acquired with the ACQ/IMAGE protocol
and found that they are well centered and isolated.
The photon distribution of the observations
in the cross-dispersion direction
is also single-peaked and  well centered
for all targets (except for \#69336, see below).

We then examined the extracted spectra of all available exposures
per target/visit and contrasted them against the co-added spectrum to 
check against variability of absolute flux levels and spectral lines.
We also checked that the weakest spectral features repeat over the different
exposures as confirmation that they are not spurious detections.


The \ion{P}{5}$\lambda\lambda$1118.0,1128.0+\ion{Si}{4}$\lambda\lambda$1122.5,1128.3,1128.4
blend is reconstructed from FPPOS-4 exposures only.
The red component can be used for analysis,
as it is seen in all FPPOS-4 exposures when several are available.
The sharp absorption seen in the bluest end of most of the final
co-added spectra is artificial.

We found that the continuum level of all targets remains constant throughout the visit.
The numerous absorption features at the expected transitions of \ion{Fe}{5}
between the geocoronal \ion{O}{1}$\lambda\lambda$1302.2-1306.0 line
and the \ion{C}{4}$\lambda\lambda$1548.2,1550.8 doublet repeat in the different exposures, 
and consequently are unlikely to be noise.
The profiles of the diagnostic lines
\ion{C}{3}$\lambda$1176, \ion{N}{5}~$\lambda\lambda$1238.8,1242.8, \ion{Si}{4}$\lambda\lambda$1393.8,1402.8,
\ion{C}{4}$\lambda\lambda$1548.2,1550.8, \ion{He}{2}$\lambda$1640.0 and \ion{N}{4}$\lambda\lambda$1718.0,1718.5
do not vary significantly over the noise levels. The only exception are targets
\#60449 and \#65426 (see below).


During these checks we detected recurrent instrumental defects 
at the same wavelengths of the same FPPOS setting for all the targets.
Even though the pipeline seems to have done a nice job removing them,
we list them here for reference:
    FPPOS-3:       1392-1396\AA~ (red wing of the blue \ion{Si}{4} absorption);
    FPPOS-1 and 4: 1172-1177\AA~ (basically overlapping with the complete \ion{C}{3} blend);
    FPPOS-1:       1230-1233\AA~ (blue wing of P~Cygni \ion{N}{5} absorption) and
    FPPOS-4:       1237-1243\AA~ (\ion{N}{5} P~Cygni transition from absorption to emission).
There is also a defect close to \ion{He}{2}$\lambda$1640.0 in FPPOS-4 observations,
but S/N is very poor in this spectral range anyway.

\begin{figure}
\centering
\includegraphics[width=\textwidth]{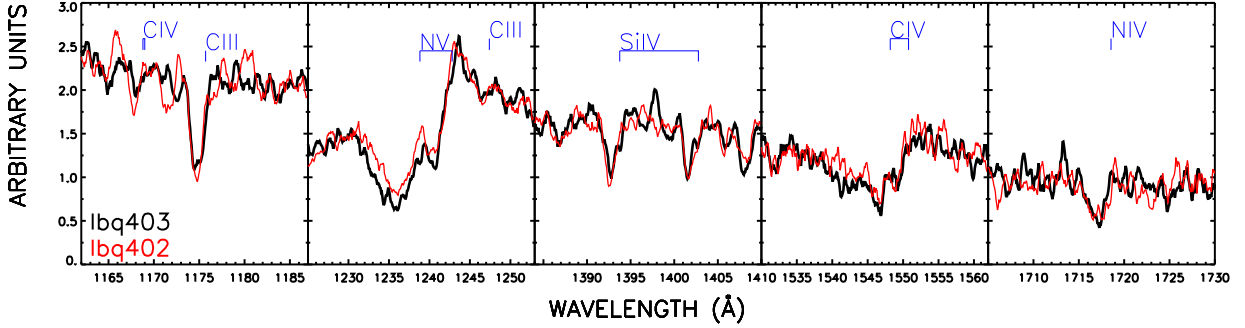}
   \caption{HST-COS observations of target \#65426, visits lbq403 (black) and 
lbq402 (red). The co-added, flux-calibrated extracted spectra of both
visits has been divided by the same arbitrary factor for clarity.
The P~Cygni profiles of \ion{C}{4} and \ion{N}{5} are more developed in visit lbq403,
which we will use for analysis.
           }
      \label{f:65426}
\end{figure}

\begin{deluxetable}{r    l     r        r            r            r        r             r     }
\tabletypesize{\small}
\tablecaption{
Summary of the input parameters and results of the SEI analysis. 
The table compiles the TLUSTY model used as incident photospheric flux
to compute the synthetic profiles.
The model was convolved by the $v_{broad}$~ velocity to
simulate the stellar rotational and macroturbulent broadening.
The observed spectrum was corrected by the radial velocity \vrad~ compiled here.
The derived wind parameters \vinf, $v_{to}$~ and $\beta$~ (see Section~\ref{s:sei}) are also provided.
   \label{T:vinf}
  }
\tablewidth{0pt}
\tablehead{
\colhead{ ID}  &   \colhead{SpT}  &   \colhead{TLUSTY}  &  
\colhead{$v_{broad}$}  &   \colhead{$v_{rad}$}  &   \colhead{\vinf}  &  
\colhead{$v_{to}$}  &   \colhead{$\beta$}  \\
\colhead{   }  &   \colhead{   }  &   \colhead{model }  &  
\colhead{[km/s]}  &   \colhead{[km/s]}  &   \colhead{[km/s]}  &  
\colhead{[km/s]}  &   \colhead{      } 
  }
\startdata 
 69217  &           O3-4Vf  &    S40000g375v10  &    94  &   -240  &   2200  $\rm ^{+  150} _{\rm -  100}$  &   150  &   0.8  \\
 62024  &         O6.5IIIf  &    S37500g375v10  &    94  &   -234  &   1250  $\rm ^{+  150} _{\rm -  200}$  &   100  &   1.2  \\
 65426  &   O7.5III-V((f))  &    S37500g375v10  &    49  &   -234  &   1500  $\rm ^{+  250} _{\rm -  250}$  &   160  &   1.4  \\
 67559  &     O8.5III((f))  &    S35000g375v10  &    50  &   -234  &   1500  $\rm ^{+  300} _{\rm -  200}$  &   130  &   0.7  \\
 63932  &             O9II  &    T35000g400v10  &   180  &   -234  &   1000  $\rm ^{+  500} _{\rm -  400}$  &    90  &   0.8  \\
 69336  &             B0Ia  &    BS25000g275v2  &   100  &   -100  &   1300  $\rm ^{+  100} _{\rm -  100}$  &   130  &   0.8  \\
 62390  &           B0.5Ia  &    BS25000g275v2  &    50  &   -255  &   1075  $\rm ^{+   75} _{\rm -   75}$  &    90  &   0.8  \\
 60449  &           B1.5Ia  &    BS22000g250v2  &    50  &   -243  &    875  $\rm ^{+   75} _{\rm -   75}$  &    90  &   0.8  \\
\enddata
\end{deluxetable}

Target \#65426 was visited twice with a separation of seven days (lbq402 and lbq403).
The individual exposures within the same visit look invariant,
however there are differences between the two visits (see Fig.~\ref{f:65426}):
the \ion{N}{5}~$\lambda\lambda$1238.8,1242.8 and \ion{C}{4}$\lambda\lambda$1548.2,1550.8 P~Cygni profiles are more developed in lbq403,
and the \ion{C}{3}$\lambda$1176 line is better defined. 
Because no continuum level variations are detected, binary effects are unlikely.
Instead, the amplitude of the changes observed in Fig.~\ref{f:65426} 
is consistent with the typical one-day timescale variability
reported for O-stars since the IUE mission \citep{PH86}.
We will use the spectrum from the lbq403 visit alone in the remainder of the paper
because of its improved S/N.


Target \#60449 also exhibits mild variability.
The blue component of the \ion{Si}{4}$\lambda\lambda$1393.8,1402.8 doublet experiences small variations.
The \ion{N}{5}~$\lambda\lambda$1238.8,1242.8 features are very weak, however present in all exposures.
They also vary, even developing to a small P~Cygni in some exposures.
For this star we detected a strong unidentified line at 1675\AA.

Finally, we detected that a second target entered the aperture
during the observations of star \#69336
from the double-peaked distribution of events of its rawtag file.
Unfortunately this target was acquired with the PEAKXD+PEAKD protocol
and there is no acquisition image available.
Because of CALCOS's fixed window for spectral extraction, 
the flux from both stars is included in the final extracted spectra.
We compared their number of events at several wavelengths, and estimated 
that the secondary target typically registers 20\% of the number
of events of the primary one.
We note that the strongest features of the composite spectrum do not significantly vary 
between different exposures which, on the other hand,
were taken within the same visit (i.e., within $\sim$3 hours).
The iron features in the pseudo-continuum do vary between different exposures
and are not reliable.

%

\subsection{Normalization}
\label{ss:nor}

Lastly, the spectra were normalized in preparation 
for the analysis with the SEI method.
Normalization of UV stellar spectra is a common practice,
useful to remove the contribution of the unknown stellar distance and radius,
and the interstellar extinction towards the object's line of sight. 
However, it adds additional errors to the final spectra to be compared,
erases any information on the star's SED
and may mask important information on the stellar metallicity (see Sect.~\ref{ss:norm}).

There are three main uncertainty sources when normalizing spectra
in the 1100-1800\AA~ range. The first one is the $\rm Ly_{\alpha}$~ absorption
due to interstellar (IS) hydrogen, whose red wing affects the local continuum
of the \ion{N}{5}~$\lambda\lambda$1238.8,1242.8 doublet.
%
%
The second uncertainty source is extinction,
whose curvature can change in the UV range
subject to additional parameters besides $R_V$~ and $E(B-V)$~ \citep[e.g.][]{FM07}.
The final and most severe uncertainty source is that there is no information on
where the stellar continuum actually is, even if extinction were negligible.
The  1200-1800\AA~ wavelength range contains numerous metallic
lines of Fe-group elements.
The lines blend and eat up the continuum in the 1200-1800\AA~ range
even for metallicities as low as 0.1\Zsun,
leaving no (or almost none) genuine continuum points.
The global absorption due to the iron forest depends not only 
on the stellar metallicity, but also on the star's
effective temperature (\Teff), gravity (\logg), rotational velocity (\vsini) and microturbulence ($\xi$).
Lacking reliable continuum points, the use of
a synthetic continuum flux as the continuum function
for normalization would be impractical;
it would require knowledge of the exact stellar radius and distance
besides the listed photospheric parameters and the details of the extinction curve
(which is not known \textit{a priori}).

We used an equivalent approach.
For each star we produced a normalized spectrum that
matches the pseudo-continuum of a normalized synthetic spectrum with similar stellar parameters.
Note that neither of them will necessarily reach unity at the pseudo-continuum.
The reference wavelengths where the observed and the model spectrum must match
were chosen from a model with similar stellar parameters to the star
and 0.1\Zsun, to avoid
artificially lowering the pseudo-continuum as much as possible.
The normalization function is a smooth polynomial calculated
by fitting the reference points with an order-3 splines function.

Besides the general uncertainty sources described earlier in this Section,
our normalization is hampered at longer wavelengths by the 
poor spectral S/N at $\lambda >$ 1600\AA.
In addition, the bluest 1100-1150\AA~ region
is affected by COS's decreased sensitivity.
Finally, the IS $\rm Ly_{\alpha}$~ absorption could not be properly modeled for our program stars
because it is greatly masked by geocoronal $\rm Ly_{\alpha}$~ emission.
We used an arbitrary unsaturated absorption 
to rectify the continuum in this spectral region.

\begin{figure}
\centering
\includegraphics[width=\textwidth]{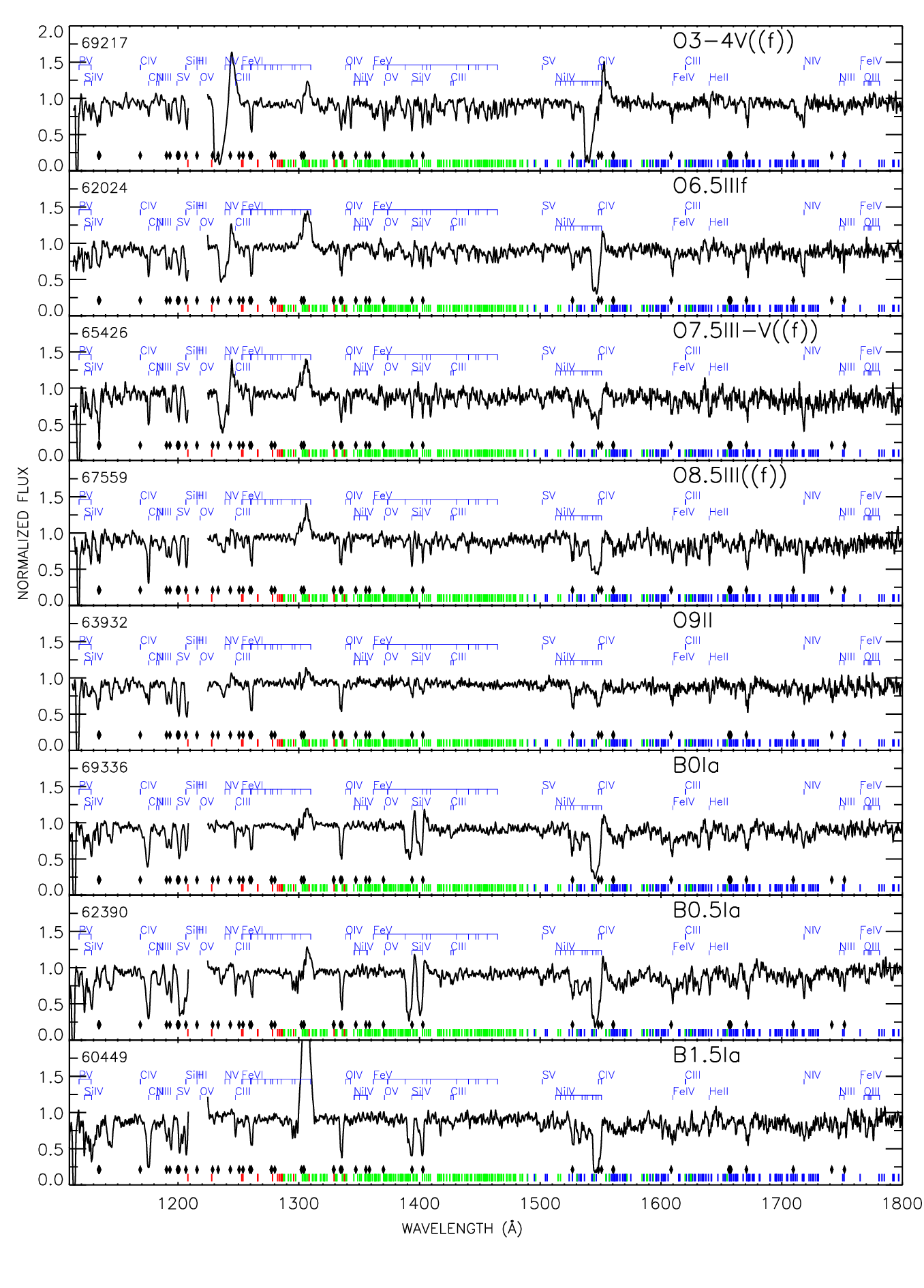}
   \vspace{-1.5cm}
   \caption{\footnotesize
HST-COS observations of O and early-B stars in IC~1613. 
The $\rm Ly_{\alpha}$~ region has been removed for clarity.
The transitions of \ion{Fe}{6} (red), \ion{Fe}{5} (green) and \ion{Fe}{4} (blue) are marked at the bottom of each panel.
Diamonds represent interstellar and air-glow lines. 
The spectra have been corrected by each star's radial velocity, and COS oversampling has been removed
for increased S/N.
           }
      \label{f:all}
\end{figure}

The final normalized spectra are shown in Fig.~\ref{f:all}.

\begin{figure}
\centering
\includegraphics[width=\textwidth]{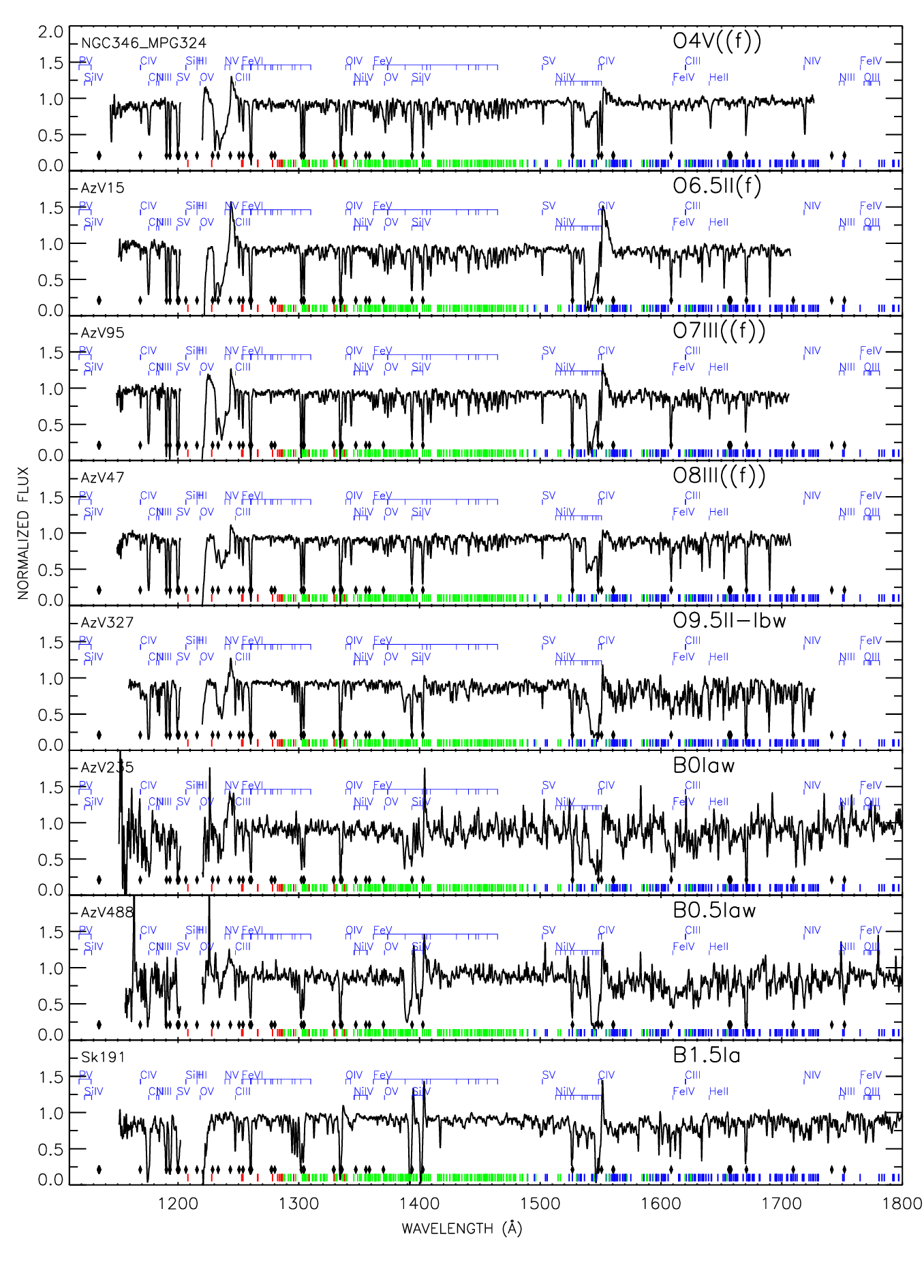}
   \vspace{-1.5cm}
   \caption{\footnotesize
UV spectra of SMC stars with spectral types matching those of the IC~1613 sample.
All spectra were taken with HST-STIS except for AzV235 and AzV488, observed with IUE.
All have been convolved with COS resolution.
           }
      \label{f:smc}
\end{figure}

\begin{figure}
\centering
\includegraphics[width=\textwidth]{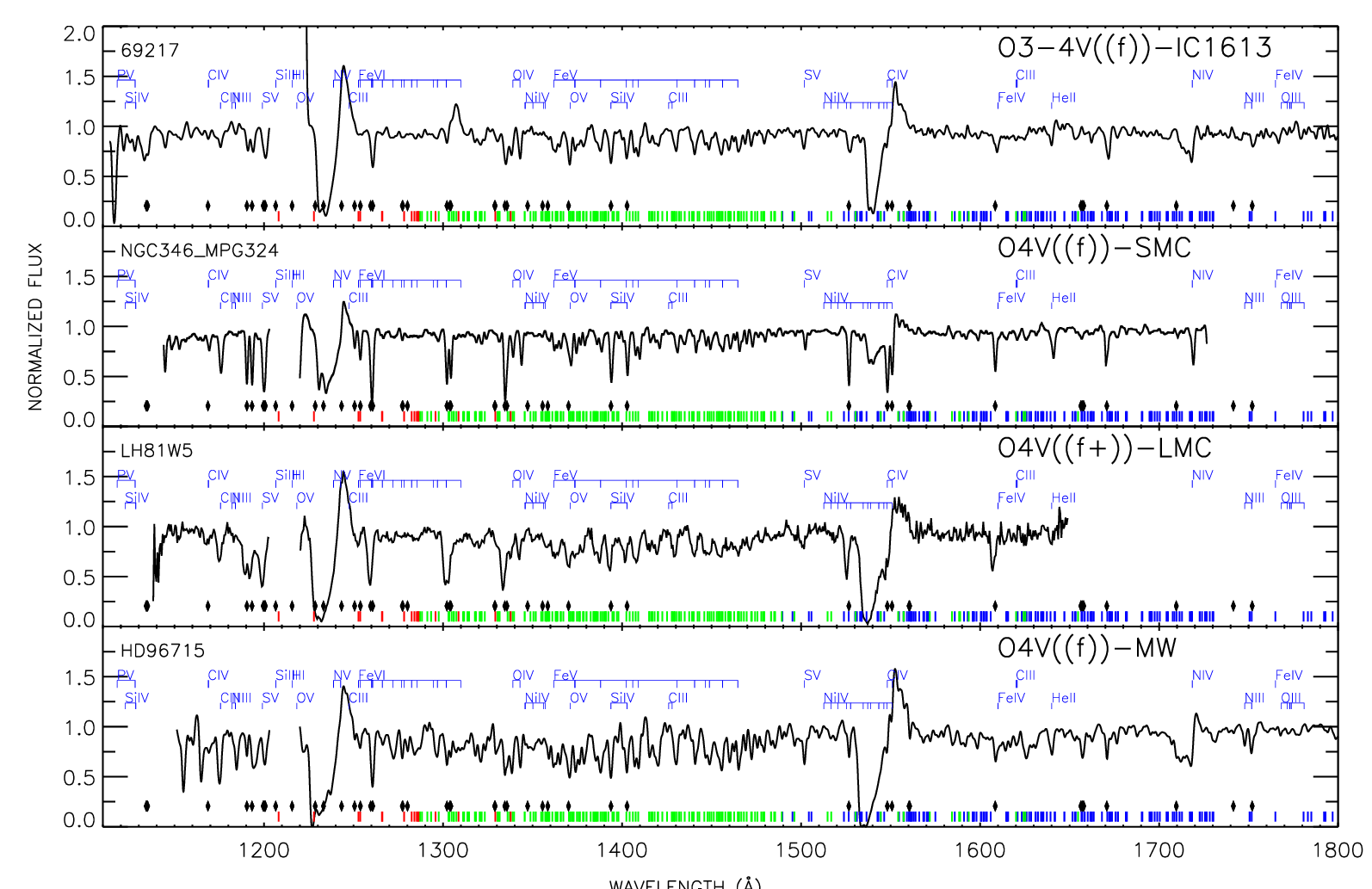}
   \caption{
Metallicity effects on the UV spectra of O3-4 type Local Group stars.
The spectrum of \#69217 (IC~1613) is compared to NGC346-MPG324 (SMC),
LH81W5 (LMC) and HD~96715 (MW).
All datasets have been degraded to match the low resolution 
of the spectra of the LMC star (HST-STIS, G140L, R$\sim$1200).
\#69217's spectral morphology is more similar to the LMC target
than to the SMC counterpart, especially in the \ion{Fe}{5}-dominated region.
           }
      \label{f:metal}
\end{figure}

\section{Spectral morphology}
\label{s:mor}

The UV spectra of the OB stars in IC~1613 are shown in Fig.~\ref{f:all}, arranged from early to later spectral type (top to bottom).
All sample stars display wind features or P~Cygni profiles of at least one highly ionized species 
(\ion{N}{5}, \ion{C}{4}, or \ion{Si}{4}). 
The strong line seen in emission in all the spectra is the \ion{O}{1}$\lambda\lambda$1302.2-1306.0 air-glow line.

Given the spectral sampling of the set of stars, we cannot discuss spectral
type effects for each luminosity class separately.
The only dwarf star of the sample is \#69217. It shows fully developed, almost saturated
P~Cygni profiles of \ion{N}{5}~$\lambda\lambda$1238.8,1242.8 and \ion{C}{4}$\lambda\lambda$1548.2,1550.8,
consistent with the high luminosity of early-O dwarfs.
%

When the overall sample of O-stars is considered, \ion{N}{5}~$\lambda\lambda$1238.8,1242.8 and \ion{C}{4}$\lambda\lambda$1548.2,1550.8
grow stronger towards earlier spectral types. 
Luminosity effects on the \ion{C}{4} doublet can also be seen,
this feature being stronger towards luminosity class-I.
\ion{C}{3}$\lambda$1176 and \ion{Si}{4}$\lambda\lambda$1393.8,1402.8 exhibit
photospheric profiles.
\ion{C}{3}$\lambda$1176 increases towards later spectral types and
remains photospheric in most early-B supergiants.
It may be a good temperature indicator for future quantitative analyses.

The sample of B-supergiant stars span only a small range of spectral types,
and no clear trend is detected.
They all show well developed \ion{C}{4}$\lambda\lambda$1548.2,1550.8  profiles,
while the \ion{N}{5}~$\lambda\lambda$1238.8,1242.8 doublet is very weak or absent.
The \ion{Si}{4}$\lambda\lambda$1393.8,1402.8 doublet displays two strong and deblended wind profiles
in all of them. \ion{C}{3}$\lambda$1176 is photospheric except in \#69336,
where the blend could be starting to develop a wind profile.

The decreased spectral quality at $\lambda >$1600\AA~ affects two diagnostic features:
\ion{N}{4}$\lambda\lambda$1718.0,1718.5 and \ion{He}{2}$\lambda$1640.
\ion{N}{4}$\lambda\lambda$1718.0,1718.5 displays a P~Cygni profile in \#69217,
but the feature is weak in the rest of the total OB-sample 
and no clear trend is detected.
\ion{He}{2}$\lambda$1640 also looks roughly constant at the spectral resolution and scale of Fig.~\ref{f:all}.
The explanation is that the \ion{He}{2} line overlaps with two iron lines \ion{Fe}{4}$\lambda\lambda$1640.0,1640.2.
The apparently constant feature-strength is likely caused by
the varying equivalent width of the individual lines with temperature, which compensates in the final
unresolved blend.

The spectral continuum of the sample O-stars is depleted 
in the 1340-1480\AA~ spectral range, which contains numerous \ion{Fe}{5} lines.
The B-supergiant stars are flat in this region, but show a depression
at $\lambda > $1520\AA, dominated by \ion{Fe}{4} transitions.
The spectrum of star \#67559 -O8.5~III((f))- exhibits intermediate features,
and  \#63932, with a quite flat spectrum, is out of the trend (see below).
This reflects the change in the ionization balance of iron as the temperature
decreases from O to B stars.

\subsubsubsection{\#63932}

We discuss target \#63932 individually, as its morphology
differs from a typical O9~II star.
Its wind profiles are weaker than expected for its luminosity class
and the absorption of the photospheric lines of
\ion{S}{5}$\lambda$1500, \ion{Si}{4}$\lambda\lambda$1393.8,1402.8 and \ion{C}{3}$\lambda$1176 
are broad.

The O9~II type was determined from our optical multi-object
spectroscopic program in IC~1613 carried out with VLT-VIMOS \citep{Gal14}.
The classification was complicated as \ion{He}{2}$\lambda$4686 could be in emission, indicating class-I,
but the broad Balmer lines and the \ion{Si}{4}$\lambda$4089/\ion{He}{1}$\lambda$4121 ratio are consistent with class III-V.
The observed stellar magnitude is also consistent
with a luminosity class between giant and supergiant, thus we adopted
luminosity class II.
Our classification agrees with \citet{Fal07}'s O9~I type,
who noted no special problem on the star.

After the HST observations were completed we learnt that \#63932 is an eclipsing binary 
discovered by the Araucaria project \citep{B13}.
This finding could explain its abnormal spectral morphology,
and indicate that the stellar magnitude could have been overestimated.
We note here that we did not detect any resolved nearby star in the acquisition image
nor any variation of the spectral lines of \#63932.
There are small variations of \ion{N}{5}~$\lambda\lambda$1238.8,1242.8,
\ion{C}{3}$\lambda$1176, and \ion{C}{4}$\lambda\lambda$1548.2,1550.8, but considering
the spectral S/N and the instrumental defects they are not significant
enough to account for binarity.
The absorptions at the Fe pseudo-continuum do vary between different exposures, and may be artificial.




%
%

\subsection{Metallicity effects on spectral morphology}
\label{ss:mm}
Our UV spectral collection is the first with 
allegedly poorer metal content than the SMC, and therefore interesting to inspect
for metallicity effects.
The archive spectra of SMC stars with matching spectral types are shown in
Fig.~\ref{f:smc}. The data were mostly taken from the high S/N, R$\sim$46000 HST-STIS observations
of programs GO7437 and GO9116 (P.I. D. Lennon\footnote{http://www.roe.ac.uk/$\sim$cje/stis.html})
and the StarCAT archive\footnote{http://casa.colorado.edu/$\sim$ayres/StarCAT/}.
The IUE INES-archive spectra (R$\sim$10000) of two targets with no HST observations were also used.
The data were normalized following the procedure explained in Section~\ref{ss:nor},
and convolved to match our COS data R$\sim$2600 resolution.

The wind profiles seen in IC~1613 spectra (Fig.~\ref{f:all}) are comparable in strength to the SMC counterparts (Fig.~\ref{f:smc}). 
However, the \ion{N}{5}~$\lambda\lambda$1238.8,1242.8 and \ion{C}{4}$\lambda\lambda$1548.2,1550.8 lines are slightly stronger
in the SMC giants and supergiants.
Whether this is reflection of the weaker winds expected for IC~1613 stars can only be determined
after quantitative analysis and will be pursued in a forthcoming paper.
In any case the lines are not saturated in the IC~1613 or SMC stars,
in contrast with \ion{C}{4}$\lambda\lambda$1548.2,1550.8 being saturated basically in all the Galactic O stars,
regardless the luminosity class \citep[e.g. ][]{BG02,GB04}.

An exception to these statements is \#69217. Its wind profiles of
\ion{N}{5}$\lambda\lambda$1238.8,1242.8 and \ion{C}{4}$\lambda\lambda$1548.2,1550.8 
are basically saturated.
They are similar to the profiles observed in the LMC and the MW (see Fig.~\ref{f:metal})
and stronger than the SMC counterpart  NGC346-MPG324. 

Neither IC~1613 nor SMC mid-O giants have wind profiles for \ion{Si}{4}$\lambda\lambda$1393.8,1402.8,
in contrast with the observed profiles for MW analogue stars. \citet{Wal00} interpret this
as a systematic effect due to low metal content.
Likewise, \ion{C}{3}$\lambda$1176 displays a P~Cygni profile in the spectra of MW mid-O giants \citep{BG02}, 
but the profile is photospheric in the analogue IC~1613 and SMC stars.

An unexpected finding is that 
when comparing analogue stars in the \ion{Fe}{5} and \ion{Fe}{4} pseudo-continua,
the line-strength of the Fe forest in IC~1613 stars has similar strength
to SMC stars, or even stronger.
We expected a significant difference given IC~1613's poorer metal content.
This is better illustrated in Fig.~\ref{f:metal},
showing the UV spectra of \#69217, NGC346-MPG324, 
the LMC star LH81W5 (HST-STIS-G140L data),
and HD96715 in the MW (IUE-SWP data).

\#69217's morphology resembles more closely the spectrum of the LMC counterpart LH81W5,
than the SMC counterpart NGC346-MPG324.
The similarity is particularly remarkable in the 1340-1480\AA~ region depleted by the numerous \ion{Fe}{5} lines.
Since there are no marked differences between the \ion{Fe}{4}- and \ion{Fe}{6}-dominated regions of the three stars, 
this fact can be discarded as a temperature effect.
Since we have no reason to suspect that the microturbulence of the three targets is different,
Fig.~\ref{f:metal} therefore suggests that \#69217's iron content is intermediate between the LMC and the SMC.

\section{SEI analysis}
\label{s:sei}

The terminal velocities of the sample stars were determined with the
Sobolev plus Exact Integration (SEI) method \citep{sei}.
In particular, we used J. Puls's implementation (private communication) of \citet{H95}'s code.
From \textit{ad hoc} input parameters and the provided photospheric flux at the base of the wind,
the code calculates synthetic spectra
using the Sobolev approximation.
The parameters are varied until the resulting profiles reproduce well the observed P~Cygnis.



The input parameters can be divided in two sets \citep[see ][]{Hal95}.
The first set, common to all the lines, includes
the turbulent velocity in the inner wind ($v_{ti}$),
the turbulent velocity in the outer wind ($v_{to}$),
the terminal velocity (\vinf) and
the wind velocity law ($\beta$).
The turbulent velocity evolves through the wind
as a function of the local wind velocity $v(r)$:
$v_{turb}(r) \, = \, v_{to} \, + \, \frac{v_{to}-v_{ti}}{1-v(r_i)} \, (v(r)-1)$.
The second set, specific to each line, 
constrains the distribution of the ion abundance 
at different layers ($k(v)$).

The parameters were derived in several iterations.
In the first iteration we used an initial \vinf~ estimate from the bluest extent of the P~Cygnis, 
we set $v_{ti}=0.02 \cdot$\vinf~
and $v_{to}=0.1 \cdot$\vinf~ \citep[following ][]{H95}
and kept $\beta=0.8$ constant.
The parameters constraining the ion stratification were then adjusted
to roughly reproduce the depth of the P~Cygni's absorption and the
transition to emission, and the location of the emission maximum.
Finally, $v_{to}$ was adjusted to match the slope of the bluest edge of
the P~Cygni absorption. The resulting profile likely requires re-adjusting \vinf,
and the process is iterated until finding a satisfactory fit.

Since unsaturated lines suffer from a small degeneracy for the [\vinf,$k(v)$] pair,
and because the terminal velocity is common to all lines,
the degeneracy is usually minimized by fitting simultaneously at least two wind profiles: 
\ion{C}{4}$\lambda\lambda$1548.2,1550.8 and \ion{N}{5}$\lambda\lambda$1238.8,1242.8, or 
\ion{C}{4}$\lambda\lambda$1548.2,1550.8 and \ion{Si}{4}$\lambda\lambda$1393.8,1402.8 depending on the star's spectral type.

None of our SEI-generated profiles reproduces well the emission of the observed  P~Cygnis.
This is partly due to 
the chosen illuminating photospheric model (Sect.~\ref{ss:tlusty}),
the uncertainty in the pseudo-continuum because of normalization (Sect.~\ref{ss:norm})
and the spectral S/N. 
For a good representation of the pseudo-continuum,
the input model should have the same Fe abundances and stellar parameters
(\Teff, \logg, \vsini, and $\xi$) as the star.
Element abundances also have an impact, as the photospheric line
affects the P~Cygni's transition from absorption to emission.

As a consequence $\beta$~ cannot be firmly constrained from our analysis,
yet this parameter may alter the overall shape of the
P~Cygni profiles and the ensuing derived \vinf. 
To address this issue we fitted each line for  
additional values of
$\beta$= 0.6, 1.0, 1.5, 2.0
in a final iteration.
Note that a similar approach was followed by \citet[][]{Eal10}.
As we expected, equally (in)satisfactory fits were found for all $\beta$,
although with  no large differences in the derived \vinf~ (see Fig.~\ref{f:norm_beta}).
The small profile differences caused by $\beta$~ variations
are absorbed by $v_{to}$, the other poorly constrained parameter of this analysis.

The error bars for the terminal velocities were derived as follows:
for the maximum and minimum considered $\beta$~values (0.6 and 2.0), we varied \vinf~
until we found that the SEI generated profile was no longer compatible
with the observations. The final error bars
enclose all compatible terminal velocities.
We comment on additional sources of uncertainty in Appendix~\ref{s:un}.
We are confident that the terminal velocity is reliable within the provided error bars.

\begin{figure}
\centering
\includegraphics[width=\textwidth]{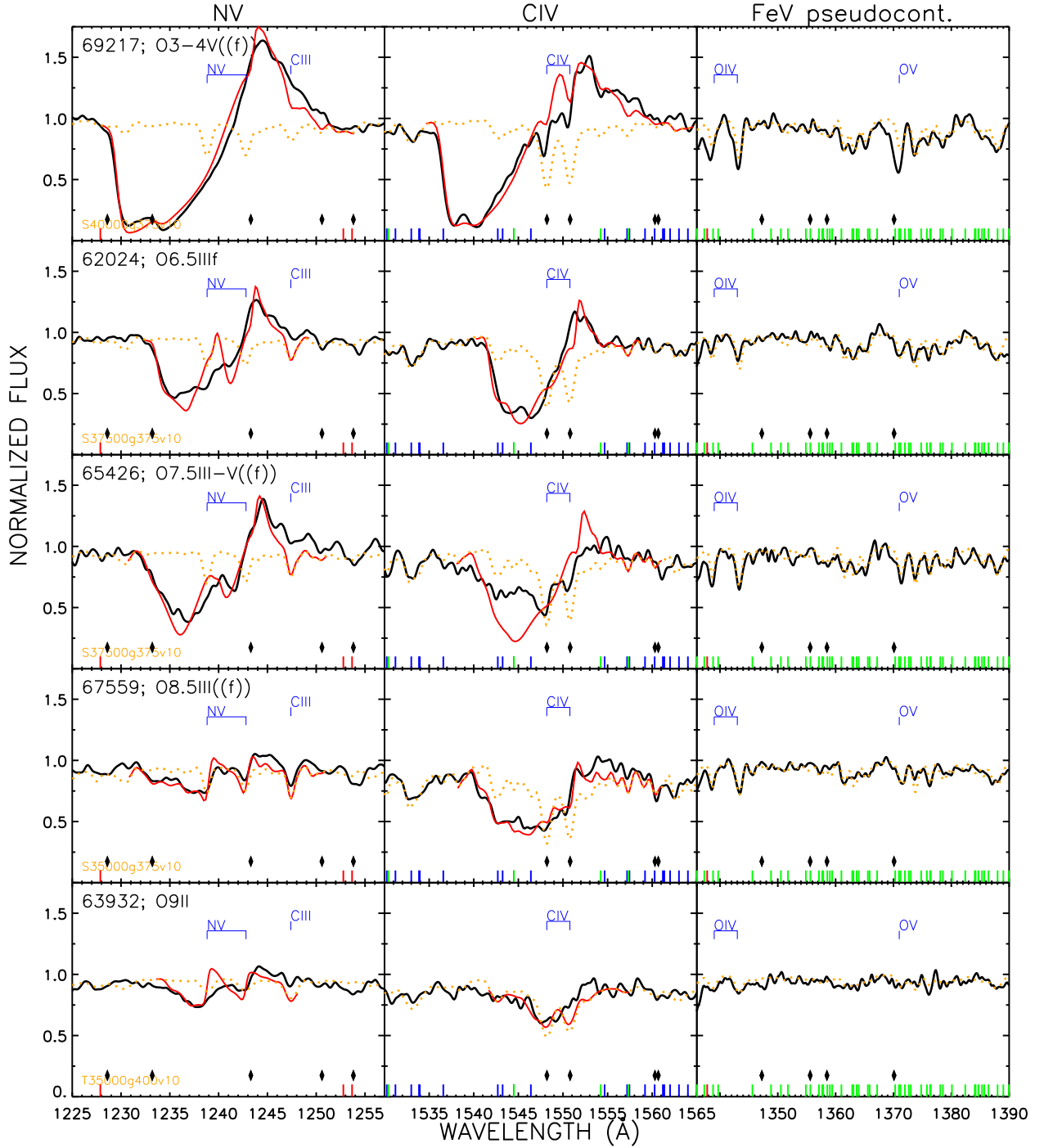}
   \caption{Best fit SEI-computed synthetic profiles (red) to the 
HST-COS spectrum (black) of the sample O stars.
The oversampling of COS's spectrum has been removed for clarity.
The TLUSTY model used as incident photospheric
flux is plotted in orange for reference. 
A portion of the pseudo-continuum is shown to
illustrate how the TLUSTY model reproduces it.
Diamonds mark interstellar lines.
The small vertical lines mark transitions of Fe.
           }
      \label{f:fitO}
\end{figure}
\begin{figure}
\centering
\includegraphics[width=\textwidth]{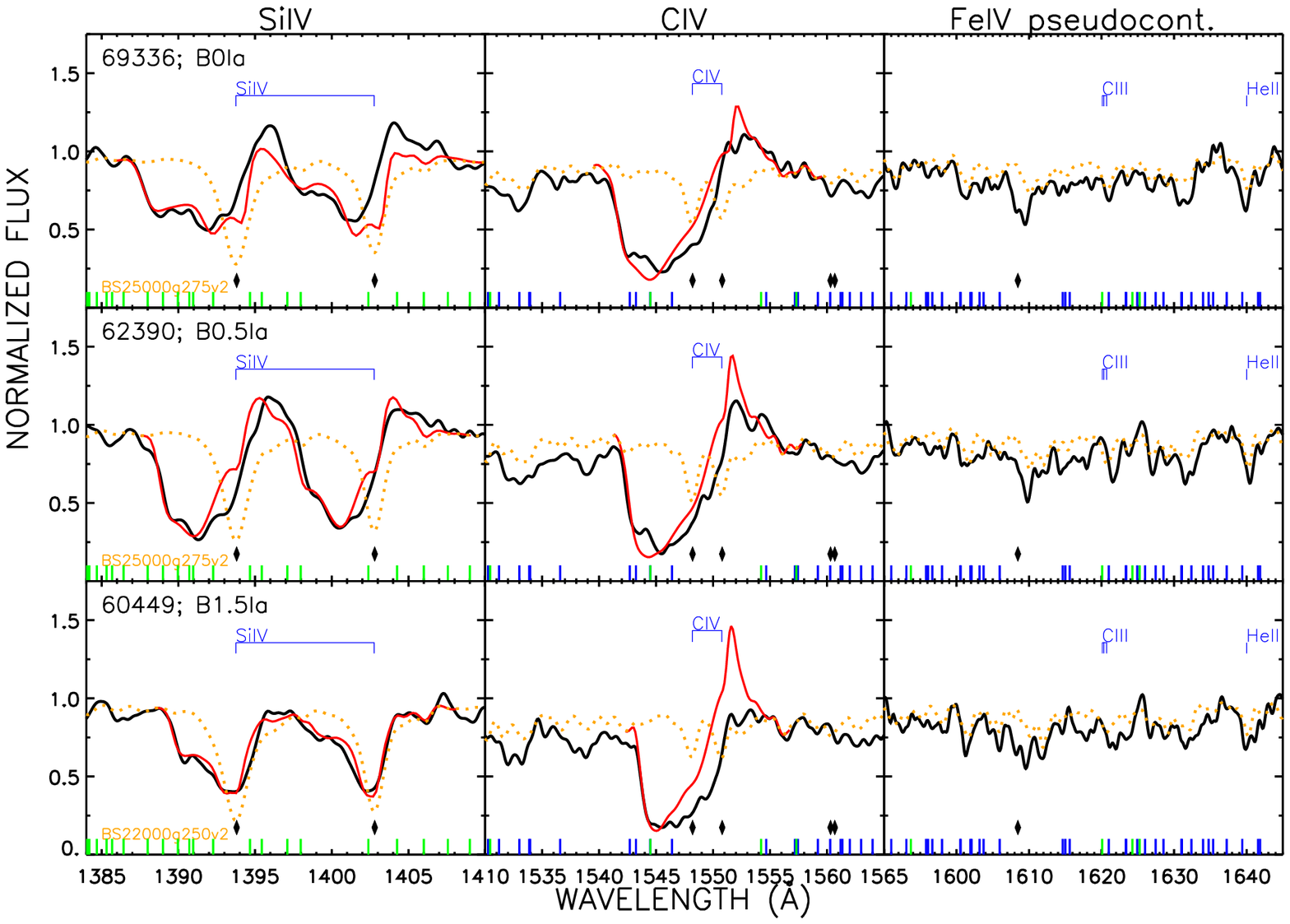}
   \caption{
Same as Figure~\ref{f:fitO}, B-supergiant sample stars.
In this temperature regime the \ion{Si}{4} P~Cygni profile
replaces \ion{N}{5} as indicator for terminal velocity, which is very weak
or absent.
           }
      \label{f:fitB}
\end{figure}

The best-fit SEI profiles are shown in Figures~\ref{f:fitO} and \ref{f:fitB},
and the derived parameters are provided in Table~\ref{T:vinf}.
Comments on individual stars are provided in Sect.~\ref{ss:com}.

\subsection{TLUSTY photospheric models}
\label{ss:tlusty}


The SEI algorithm requires that the user provides
the incident flux in the inner wind.
A realistic illuminating photospheric flux
that reproduces well the Fe pseudo-continuum is necessary
to fit the P~Cygni emission. 
Otherwise, all the SEI generated profiles have excess emission,
regardless of $\beta$.
Additionally, the lines of our sample spectra are not saturated and show their
photospheric components.
The photospheric lines must be taken into account to properly model the P~Cygni's transition
from absorption to emission.

In the past the use of realistic input photospheric fluxes was given small consideration
since these were considered only aesthetic effects with no impact on the resulting terminal velocities
\citep[e.g.][]{Mal04,Eal04c}.
However, we have detected iron lines of non-negligible equivalent width
that overlap with the blue edge of the \ion{C}{4}$\lambda\lambda$1548.2,1550.8 P~Cygni.
They mostly belong to \ion{Fe}{4} and grow stronger with decreasing temperature.
Several examples are shown in Sect.~\ref{ss:com}.
\textit{If unaccounted for, we caution that these lines may hamper the determination of the terminal velocities
of the latest O-types when \vinf$\sim$1200\kms}.

We therefore used TLUSTY models with appropriate stellar parameters for each target
as input for the SEI routine.
TLUSTY \citep{HL95} calculates plane-parallel, line-blanketed  atmospheres in hydrostatic equilibrium,
allowing for departures from local thermodynamic equilibrium (LTE).
It can compute a wealth of non-LTE transitions of metallic elements,
and is therefore accurate to model the photospheric lines of UV spectra.
We used the models from \citet{LH03}'s grid\footnote{From http://nova.astro.umd.edu/Tlusty2002/tlusty-frames-OS02.html}.
It covers the [\teff,\logg] space parameter for OB stars in 2500~K steps for effective
temperature, and 0.25 for gravity,
with $\xi=$10\kms~ microturbulence for O-type stars and $\xi=$2\kms~ for the B-stars.

For each star we used TLUSTY models with
similar stellar parameters to those derived for the star by
previous works \citep{Fal07,Tal11,Hal12,Grecia,Gal14}.
We used models from the libraries with metallicity 1/2\Zsun, 1/5\Zsun~ and 1/10\Zsun,
corresponding to the LMC, the SMC and IC~1613.
They were convolved to simulate the rotational (\vsini)
broadening of the star.
Macroturbulent broadening measurements ($\Theta_T$) are also available
for the targets that overlap with our VIMOS sample \citep{Gal14}.
For these, the total applied broadening was calculated from the
quadratic sum of \vsini~ and $\Theta_T$.
Finally, the models were convolved to match the 0.48\AA~
resolution of the observations,
and the observations were corrected from radial velocity \vrad.

The normalized synthetic and observed spectra were contrasted
at the pseudo-continuum between \ion{N}{5}$\lambda\lambda$1238.8,1242.8 and \ion{O}{1}$\lambda\lambda$1302.2-1306.0 
(dominated by \ion{Fe}{6} lines) and
between \ion{O}{1}$\lambda\lambda$1302.2-1306.0 and \ion{C}{4}$\lambda\lambda$1548.2,1550.8 (\ion{Fe}{5} lines).
The region between \ion{C}{4}$\lambda\lambda$1548.2,1550.8 and \ion{N}{4}$\lambda\lambda$1718.0,1718.5 provides information
on the \ion{Fe}{4} transitions, but S/N is poor at these wavelengths.
Several models with varying \Teff, \logg and Z were evaluated,
and the one that best reproduced the relative strengths of
\ion{Fe}{4}, \ion{Fe}{5} and \ion{Fe}{6} lines was used as input for the SEI code.

Note that the forest of Fe lines suffers from the usual [\teff,\logg] interdependence,
but there are also  Z-\logg~ and Z-$\xi$~ degeneracies.
By increasing gravity (with increasing Stark broadening) or microturbulence
the myriad of Fe lines grow stronger, the overlap increases
and the resulting pseudo-continuum is lower,
mimicking the effect of increased metallicity.
Therefore our selection of a TLUSTY model must not be taken as an exact determination
of the stellar parameters, which requires tailored analysis of additional diagnostics,
but as finding a good representation of the continuum around the lines to be analyzed 
with SEI.

The selected TLUSTY models are registered in Table~\ref{T:vinf}.
For most of the sample stars we find that the models with SMC metallicity
represent better the Fe pseudo-continuum than the models with 1/10\Zsun.
Even though this is not a robust metallicity determination,
these findings cast some doubts on the alleged 0.13\Zsun~ metallicity of 
IC~1613's OB stars (see Sect.~\ref{s:metal}).

\subsection{Comments on individual targets:}
\label{ss:com}

\subsubsection{\#69217:}
\#69217 is the hottest and most luminous star of the sample,
and displays the strongest P~Cygni profiles.
The SEI profiles with \vinf=2200\kms~ and $\beta$=0.8 reproduce well the
\ion{C}{4}$\lambda\lambda$1548.2,1550.8 and \ion{N}{5}$\lambda\lambda$1238.8,1242.8
P~Cygnis.
The observed photospheric component overlapped to the emission
of the \ion{C}{4}'s P~Cygni is also reproduced 
-although not totally- by the synthetic profile.

The TLUSTY models that best represent  the pseudo-continuum at the
wavelength ranges dominated by \ion{Fe}{4}, \ion{Fe}{5} and \ion{Fe}{6}
have SMC metallicity and \Teff=40000~K.
Note that this is a lower temperature than the 42800~K we derived from
the CMFGEN fitting of the joint UV and optical spectra  \citep{Grecia}. 
The \ion{C}{3}$\lambda$1247.5 line overlapped to the \ion{N}{5} emission of the SEI profile is too strong 
and it actually suggests an intermediate temperature, but we are limited by the grid coverage.
This star was also analyzed by \citet{Tal11} who obtained \Teff=47.6$^{+4.73}_{-4.95}$~kK and \logg=3.73$^{+0.13}_{-0.22}$,
with large error bars in temperature because of the weak \ion{He}{1}$\lambda$4471 line.
Tests with the WM-Basic \citep{wmbasic} and CMFGEN codes \citep{cmfgen1},
have shown that models with such increased temperature cannot
reproduce the UV Fe forest.

The spectrum is mostly consistent with the -240\kms~ radial velocity reported
by \citet{Fal07} and \citet{Tal11},
but \ion{O}{5}$\lambda$1371.0, \ion{S}{5}$\lambda$1500, \ion{Si}{4}$\lambda\lambda$1393.8,1402.8,
the \ion{C}{4}$\lambda\lambda$1548.2,1550.8 photospheric components and some
Fe features indicate \vrad=-270\kms. 
Since we encounter similar problems in other target stars
and considering the reported problems of wavelength accuracy (see Sect.~\ref{s:red}),
we believe this is a wavelength calibration problem rather
than some binarity issue.

\subsubsection{\#62024:}
\label{sss:62024}
This star exhibits unsaturated but strong P~Cygni profiles of 
\ion{N}{5}$\lambda\lambda$1238.8,1242.8 and \ion{C}{4}$\lambda\lambda$1548.2,1550.8,
which we use to derive \vinf=1250\kms~ and $\beta$=1.2.

We did not achieve a good fit for the \ion{N}{5} doublet.
All the SEI generated profiles exhibited a double absorption,
in disagreement with the observed single absorption.
A single synthetic \ion{N}{5} absorption could be produced by increasing $v_{ti}$~ to 
values close to $v_{to}$, but this is
contrary to the current working paradigm of turbulent velocity increasing from
the photosphere to the outer wind layers \citep{Hal95}.
Besides, these models would not produce any \ion{N}{5} emission, even for the largest values of $\beta$.

During the analysis, we detected a blend of photospheric \ion{Fe}{4} lines at $\sim$1542.7\AA~
in the TLUSTY model, that overlaps with the \ion{C}{4} P~Cygni bluest extent
(see Fig.~\ref{f:fitO}).
Its intensity is very small compared with the
wind absorption and in this case it does not play a role.
However, it may affect the derived terminal velocities of later O-stars with
\vinf$\sim$1200\kms.

We had previously analyzed the optical spectrum of \#62024
in \citet{Hal12}.
The data exhibit a small P~Cygni profile of \ion{He}{2}$\lambda$4686,
which we used to determine
\vinf=1800 $\pm$ 600 \kms~ and $\beta$=2.0.
The SEI analysis of UV lines has determined different parameters:
\vinf=1250 $\rm ^{+150}_{-200}$\kms~  and $\beta$=1.2.
While $\beta$=2 profiles could fit the observed \ion{N}{5} line, they are not compatible with the observed \ion{C}{4} emission
(see Fig.~\ref{f:norm_beta}).
Likewise, $\beta$=1.2 cannot reproduce the observed morphology of the \ion{He}{2}$\lambda$4686 optical line.
Note, however, that the UV terminal velocity is independent of $\beta$ 
within the provided error bars (Fig.~\ref{f:norm_beta}).
We therefore confirm the optical lower-limit of  \vinf~
for \#62024, and revise its value downwards by 550\kms.
This comparison
remarks how UV spectroscopy is crucial to properly model the winds of OB stars.

For the analysis of the optical spectral lines we used
rotational (\vsini=80\kms) and macroturbulent ($\Theta_T$=50\kms) broadening,
which we also apply to the reference TLUSTY model ($v_{broad}$= 94\kms).
Most importantly, we found no radial velocity variations
over an extended period of more than 4 weeks in the VIMOS dataset, with \vrad=-234\kms.
However, some UV photospheric lines including \ion{C}{3}$\lambda$1176 and \ion{O}{4}$\lambda\lambda$1339.0,1343.0
rather indicate \vrad=-280\kms, and this correction 
would also fit better the emission of \ion{C}{4}.
Since the star's radial velocity has been proven constant,
the discrepancy again hints a problem with the wavelength calibration.

\subsubsection{\#65426:}

The spectrum of \#65426 shows a strong, well developed P~Cygni of \ion{N}{5}$\lambda\lambda$1238.8,1242.8
and a mild profile of \ion{C}{4}$\lambda\lambda$1548.2,1550.8.

None of the SEI generated profiles reproduced the observed \ion{C}{4} morphology,
and the wind parameters were mainly derived from \ion{N}{5}.
The observations exhibit strong absorption features at the rest-frame \ion{C}{4} wavelengths, which
we suspect could be partly interstellar.
Note that the \ion{Fe}{4} absorption at $\sim$1542.7\AA~ is present in this star,
now with comparable strength to the wind absorption of \ion{C}{4}.



\subsubsection{\#67559:}

The wind features of \#67559 are weak.
The small P~Cygni of \ion{N}{5}$\lambda\lambda$1238.8,1242.8 is real, as endorsed by 
its detection in all the individual exposures of this target's visit.
However, it offered no constraints on \vinf~ and we focused on reproducing the
\ion{C}{4}$\lambda\lambda$1548.2,1550.8 profile.
%

The determination of the stellar terminal velocity 
is hampered by the Fe absorptions overlapped to the 
\ion{C}{4} wind absorption trough. 
Besides the \ion{Fe}{4} feature at $\sim$1542.7\AA~ reported in Sect.~\ref{sss:62024},
another blend of \ion{Fe}{4} lines is seen at $\sim$1540.5\AA~
in the TLUSTY model (see Fig.~\ref{f:fitO}).
There is a small interval between these lines at $\sim$1541\AA~ which seems free of photospheric 
absorption and used as reference to constrain \vinf, but we lack any information on the slope of the
blue edge of the absorption and we have no constraints on $v_{to}$.
To determine the upper \vinf~ limit for this star, the terminal velocity was increased
until the resulting P~Cygni extended bluewards of all nearby photospheric
absorptions (1540.5\AA~ for \ion{C}{4} and 1233.5\AA~ for \ion{N}{5}).

All models fail to fit the extended emission of \ion{C}{4} and \ion{N}{5}, regardless of the $\beta$~
value used. The final $\beta$=0.7 offers the best compromising fit to both.

\subsubsection{\#63932:}
As we explained in Sect.~\ref{s:mor}
this star has been recently reported an eclipsing binary \citep{B13}.
Even though we expect minimal contamination by the secondary
in this range, 
we have found that its UV spectral morphology is abnormal.

\ion{N}{5}$\lambda\lambda$1238.8,1242.8 displays the only clear P~Cygni of the spectrum
and we used it to constrain the wind properties. 
However, all the SEI calculated spectra exhibit a double \ion{N}{5} P~Cygni in disagreement
with the single emission of the observations.
The weak wind trough of \ion{C}{4} exhibits an
extended absorption from 1545\AA~ to 1540\AA,
consistent with the derived terminal velocity from \ion{N}{5}.
There are no constraints on $v_{to}$~ and $v_{ti}$, which we set
to their default values.

\#63932 is the only sample star whose spectrum is better reproduced
by a 1/10\Zsun~ TLUSTY model.
This is not conclusive evidence, however, since the absorptions
in the Fe forest vary between different exposures (Sect.~\ref{s:mor}).

We derived \vsini=150\kms~ and $\Theta_T$=100\kms~ 
from  our optical spectroscopic analysis \citep{Gal14}.
The wide lines likely reflect the binary nature of the system
but these values are nonetheless useful to simulate the observed line broadening.
An equivalent $v_{broad}$=180\kms~ was applied
to the TLUSTY model used in the SEI analysis.


\subsubsection{\#69336:}

\#69336 is the earliest B-supergiant star of the sample
and has fully developed P~Cygni profiles of \ion{Si}{4}$\lambda\lambda$1393.8,1402.8
and \ion{C}{4}$\lambda\lambda$1548.2,1550.8. Our models show that 
\ion{N}{5}$\lambda\lambda$1238.8,1242.8 still experiences wind contamination, but
it is too weak for analysis. Similarly, the core and width of the \ion{C}{3}$\lambda$1176 blend
indicates an incipient wind profile.

We managed a good fit to \ion{Si}{4} and \ion{C}{4} except for the P~Cygni transitions from absorption
to emission which are in defect in the \ion{Si}{4} profiles and in excess in the \ion{C}{4} profiles.

The radial velocity derived from the current UV dataset (-100\kms) 
is in marked discrepancy with \citet{Fal07}'s quoted value of \vrad=-250\kms.
This star was included in our VIMOS series
and the averaged spectrum indicates a similar  \vrad=-234\kms.
To further complicate the interpretation of the available data,
there is a nearby star revealed only by COS enhanced spatial resolution (see Sect.~\ref{ss:var}).
The relative shift between the -100\kms~ and -250\kms~ radial velocities is 1\AA~ at 1500\AA,
hence under the setup resolution,
but clearly this target needs follow-up with higher resolution spectroscopy.


\subsubsection{\#62390:}

The spectrum of \#62390 exhibits strong P~Cygni profiles of \ion{Si}{4}$\lambda\lambda$1393.8,1402.8
and \ion{C}{4}$\lambda\lambda$1548.2,1550.8.
\ion{N}{5}$\lambda\lambda$1238.8,1242.8 also displays a small P~Cygni, but it is 
too weak to provide useful constraints on \vinf.

We derived the wind parameters from the \ion{Si}{4} doublet because
the blue edge of the \ion{C}{4} absorption is contaminated by the 
\ion{Fe}{4} features and because all the synthetic profiles produced excess emission of
this feature.

We found again small wavelength discrepancies between the observations and the models.
Some lines, including \ion{C}{3}$\lambda$1247.5, are better matched with a radial velocity correction of \vrad=-160\kms,
but most others with \citet{Fal07}'s \vrad=-255\kms.
There is no clear trend for the Fe features, where S/N is poor,
and the wavelength discrepancy hampered the election of a TLUSTY model.
We adopted \vrad=-255\kms~ and SMC metallicity.
%

We used a low \vsini=50\kms~ for this star,
as \citet{Fal07} found that the rotational broadening was under the
resolution of their instrumental set-up.

\subsubsection{\#60449:}
\ion{C}{4}$\lambda\lambda$1548.2,1550.8 still shows a small P~Cygni 
in the COS spectrum of the latest B-supergiant star of the sample, although with
very weak emission.
The \ion{Si}{4}$\lambda\lambda$1393.8,1402.8 doublet is dominated by its photospheric components, 
but it also displays weak wind absorption troughs.

The terminal velocity was derived from \ion{C}{4}, but the synthetic spectrum
also reproduces well the weak \ion{Si}{4} wind features.
However, all the SEI profiles produce an excess emission of \ion{C}{4} and none can reproduce the transition
from absorption to emission. 
Note that the latter problem also affects other sample stars.
It could probably be remedied by using photospheric models
with varying C abundance (hence varying line-strength of the \ion{C}{4} photospheric profile),
or by taking into account the contribution of the interstellar absorption lines.
Both possibilities will be taken into account in our subsequent
follow-up analysis.

The spectrum of \#60449 also shows small radial velocity inconsistencies.
\citet{Fal07} determined \vrad=-243\kms~ which we adopt,
but other lines better match \vrad=-200\kms.
Together with the low spectral S/N, the \vrad~ problem
hampered the selection of the best fitting metallicity.
We used models with SMC metallicity broadened
by \vsini=50\kms, similarly to \#62390.

\begin{figure}
\centering
\includegraphics[]{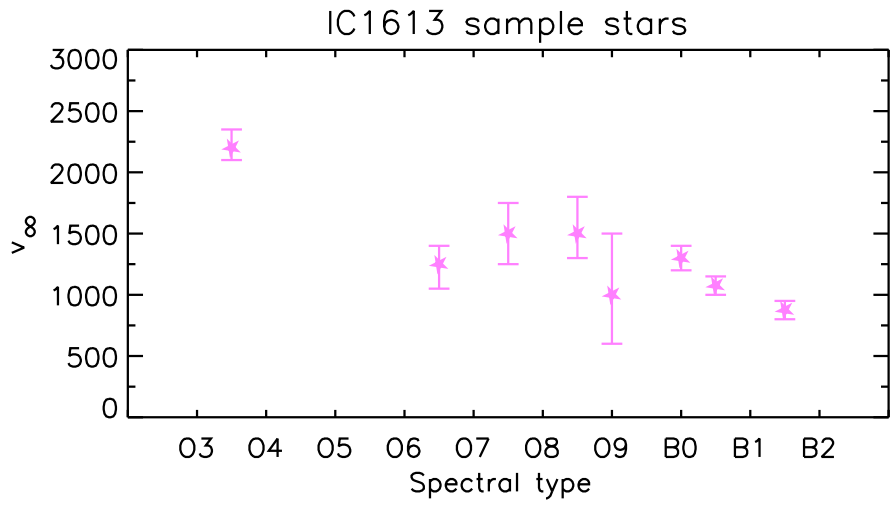}
   \caption{Terminal velocities derived in this work for 
the sample of OB-stars in IC~1613.
           }
      \label{f:vi}
\end{figure}

\section{Results}
\label{s:r}


The terminal velocities derived for the sample stars are shown in 
Fig.~\ref{f:vi} as a function of spectral type.
The overall sample follows a trend of decreasing terminal velocity
towards later spectral types with some scatter.
The behavior of \vinf~ as a function of more meaningful stellar
parameters and in the context of the Local Group
is discussed in Sect.~\ref{s:vinfZ}.

One of the main motivations for this work was to obtain an improved WLR
for IC~1613,
in sight of the existing reports of member stars 
with stronger wind momentum than predicted by theory.
%
We have used the \logq values derived for \#62024 and \#69217 from optical studies 
and our \vinf's to recalculate their mass loss rates and wind momentum.
The results illustrate how
important accurate terminal velocities are when building and interpreting the WLR.

\begin{figure}
\centering
\includegraphics[width=\textwidth]{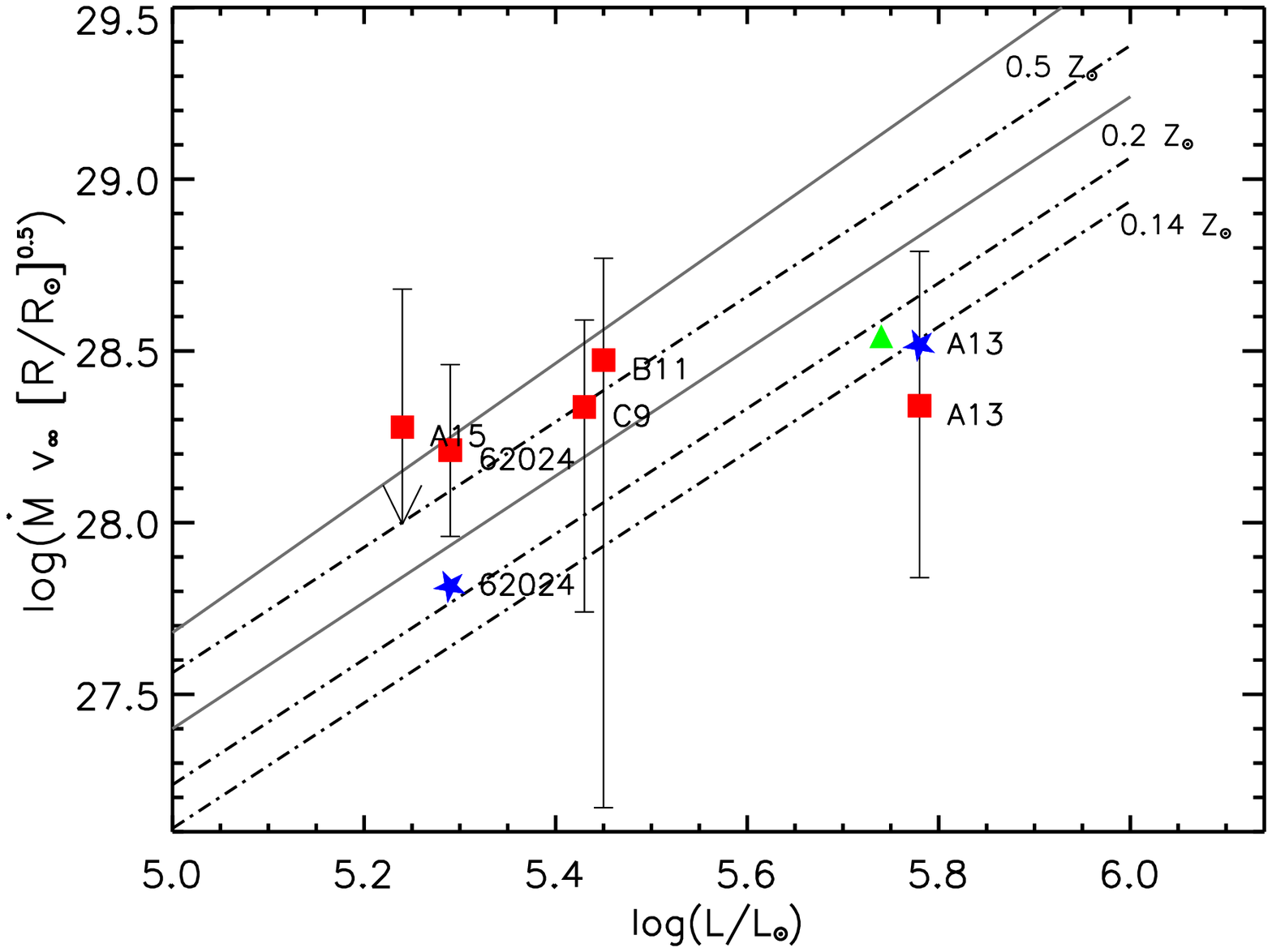}
   \caption{IC~1613's wind-momentum luminosity relation, revisited.
Squares mark wind momentum values calculated with the wind parameters derived by \citet{Tal11}
and \citet{Hal12}. Stars mark the updated wind momentum for the two targets whose terminal
velocity has been measured in this paper: \#62024 and \#69217 \citep[A13 in][]{Fal07}.
The triangle marks our results for \#69217 from a joint UV+optical analysis \citep{Grecia}.
\citet{VKL01}'s predictions for the WLR at 0.14\Zsun, 0.2\Zsun~ and 0.5\Zsun~ 
are 
represented by dashed-dotted lines.
The solid lines are the empirical relations derived by \citet{Mal07b} 
from LMC and SMC studies that do not take clumping into account.
           }
      \label{f:wlr}
\end{figure}

Fig.~\ref{f:wlr} shows the wind momentum of IC~1613 stars calculated from optical spectroscopy only
\citep[][]{Tal11,Hal10,Hal12}.
Given their large error bars, these values are compatible with 
\citet{VKL01}'s prediction for IC~1613 metal content.
However, it was the systematic offset of most of the sample stars towards the
theoretical expectation for the LMC that roused alarm.

 \clearpage

Because the optical analyses did not take clumping into account, we also compare 
their results with the empirical relations derived by  \citet{Mal07b}
from the available studies on LMC and SMC targets that similarly neglected clumping.
IC~1613 targets now lie between the locii of the LMC and the SMC, partly 
diminishing the discrepancy.

\#69217 \citep[A13 in][]{Fal07} was the only star in \citet{Tal11}'s sample under the theoretical 0.14\Zsun~ WLR.
The stellar wind momentum calculated with our derived \vinf~ better matches the prediction for 
0.14\Zsun~ stars.
However, \citet{Tal11} used \halpha~ for the analysis, quite insensitive to $Q$ when the mass
loss rate is small. 
Our preliminary analysis \citep{Grecia} that included UV lines, much more sensitive to the wind mass loss,
yields a slightly stronger $D_{mom}$~ (Fig.~\ref{f:wlr}).

The updated $D_{mom}$~ value for \#62024's is lower 
than we had calculated in \citet{Hal12}.
It is consistent with \citet{VKL01}'s prediction for 0.2\Zsun~ stars, 
and would match \citet{Mal07b}'s unclumped WLR relation scaled for 0.14\Zsun.

However, these wind-momentum values are provisional.
The WLR in IC~1613 can only be properly assessed once
the mass loss rate and
the metal content of the stars (see Sect.~\ref{s:metal}),
besides our terminal velocities, are well constrained.
The calculated wind momentum
of IC~1613 stars can also decrease once
the clumping factor is determined \citep{L12}.
A detailed spectroscopic analysis of UV and optical diagnostics
that will provide accurate photospheric properties, wind parameters and abundances
is planned as future work.


\section{The metallicity dependence of the terminal velocity}
\label{s:vinfZ}

\begin{figure}
\centering
\vspace{-1cm}
\includegraphics[height=\textheight]{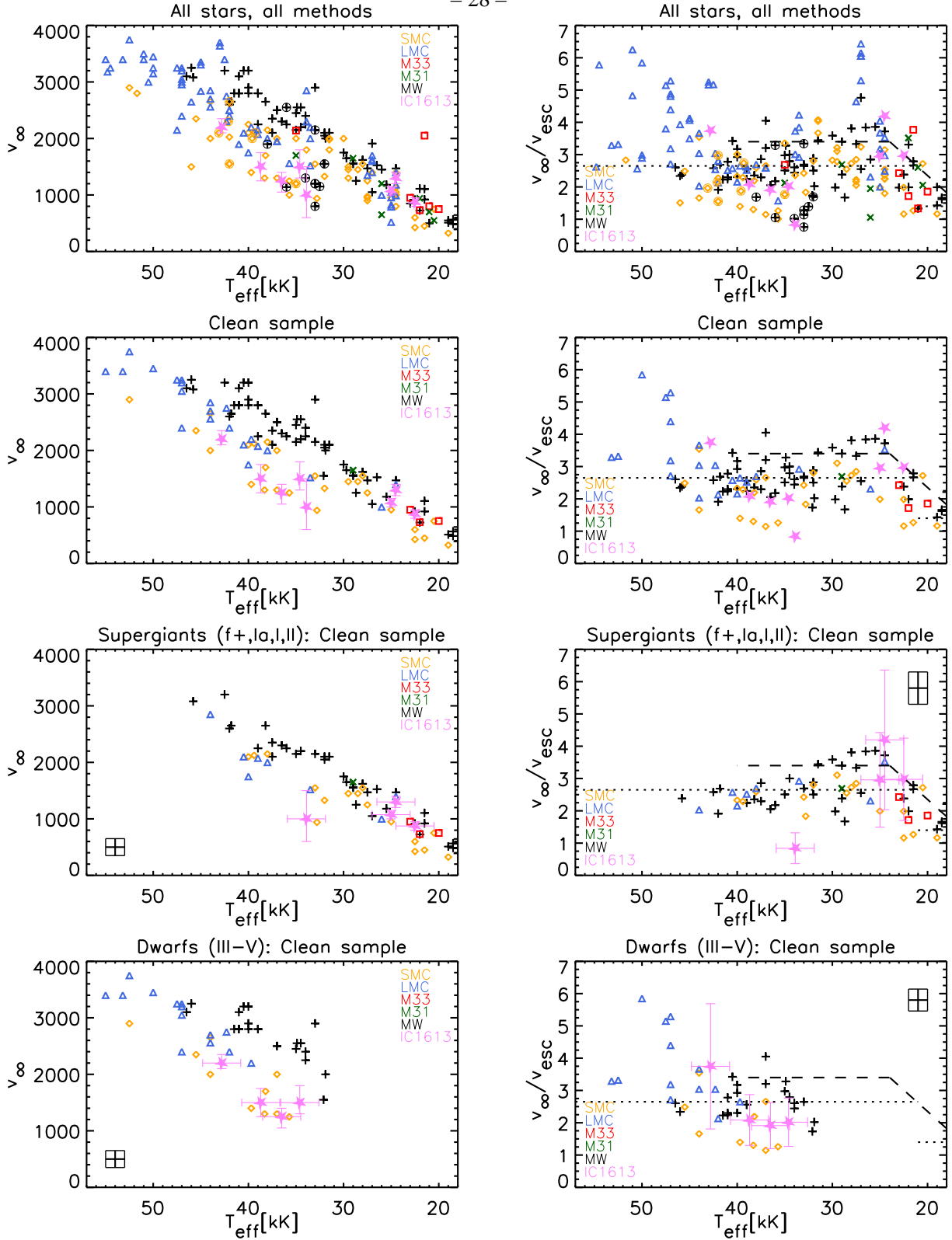}  
   \vspace{-0.5cm}
   \caption{\footnotesize
Left: Terminal velocities of Local Group O- and early-B stars
as a function of effective temperature (rhombus: SMC; triangles: LMC; 
squares: M33; crosses: M31; plus-signs: MW; stars: IC~1613).
Encircled targets have been reported to have weak winds.
The top plot shows all \vinf~ values compiled from the literature,
and the lower three rows only the cleaned sample (see text).
Right: \vinf/\vesc~ ratio of Local Group O- and early-B stars
as a function of effective temperature, same structure as the left part of the figure.
The dotted lines mark the widely used \vinf/\vesc=2.65 relation;
the dashed lines mark the \vinf/\vesc~ relations and regimes found 
by \citet{Cal06} and \citet{MP08} for Galactic B-supergiants.
Both: The framed error-bars in the lower two rows indicate typical error bars 
for the non-IC~1613, non-MW targets of the cleaned sample.
          }
      \label{f:vinfZ}
\end{figure}

Because they are costly in observing time, extragalactic O- and early-B type stars
are often observed only in the optical range. Consequently, the analyses have no information 
on the terminal velocity of the stars. \vinf~ is then taken from MW calibrations of
the terminal velocity with spectral type or effective temperature 
or estimated from the escape velocity (see for instance \citet{KP00}), 
and then scaled by \citet{LRD92}'s relation. In this section we check on the 
metallicity dependence of the terminal velocity of OB stars.

%

We have compiled from the literature the terminal velocities derived for Local Group OB stars from
UV spectroscopy (IUE, HST-STIS, HST-FOS, HST-GHRS) by
\citet{P87},    
\citet{Bal94},  
\citet{H95},    
\citet{Hal95,Hal98},  
\citet{PC98},   
\citet{Sal01},  
\citet{Bal02},  
\citet{Cal02},  
\citet{Ual02},  
\citet{Mal03},  
\citet{Hal03},  
\citet{Eal04a,Eal04c}, 
\citet{PMal04,PMal05}, 
\citet{Eal10}, 
\citet{Ral12a,Ral12b}, 
and
\citet{Bal13}.  
The sample includes targets from the SMC, the LMC, M31 and M33.
All reliable \vinf~ UV-determinations from the literature were used to increase the sample size, after checking that the inclusion of 
results from works that do not use the SEI method do not add additional scattering.

Effective temperatures, gravities and radii 
derived from quantitative spectroscopic analysis of the stars with FASTWIND or CMFGEN models
were also compiled
\citep{Hal95,  
       Pal96,  
       PC98,   
       Sal01,  
       Bal02,  
       Cal02,  
       Ual02,  
       Hal03,  
       Wal04,  
       Tal04,  
       Eal04b,Eal04c, 
       PMal04,PMal05, 
       TL05,  
       Ual05,  
       Mal06,  
       Eal10, 
       Ral12a,Ral12b, 
       Bal13} 
except for some M31 targets for which we took the parameters from \citet{Tal02}'s LTE analysis.  
For comparison, we used data on Milky Way stars from
\citet{Cal14}'s compilation (O-stars) and \citet{Cal06} results (B-supergiants).

We only kept the most recent reference per target.
Since an important fraction of CMFGEN studies adopt \logg~ and
keep it constant during the fitting process, FASTWIND analyses were preferred.
Targets with no photospheric parameters available in the literature were assigned the photospheric parameters of
a star with similar spectral type, luminosity class and metallicity.

The compiled photospheric parameters were used to calculate escape velocities,
$v_{esc} \, = \, \sqrt{2 g R_{\ast} (1-\Gamma)} \, $.
To compute $\Gamma$, which quantifies the decrease of the gravitational potential
due to Thompson scattering,
we adopted $N(He)/N(H)=0.1$ and
0.2\Zsun~ metallicity for IC~1613 and SMC stars,
0.4\Zsun~ for the LMC,
0.5\Zsun~ for M33,
1\Zsun~ for the MW
and 2\Zsun~ for M31.
\citet{Tal02} provided \Teff~ but not \Rstar~ of most of their sample; these targets were assigned the escape velocity of a star with similar
spectral type.

In Fig.~\ref{f:vinfZ} we analyze the behavior of the terminal velocity and the \vinf/\vesc~ ratio
as a function of effective temperature and metallicity.
The top row of plots shows the total compiled list of Local Group OB stars.
For the three remaining rows, the sample has been cleaned from composite, variable or peculiar targets,
or showing WN spectral features. 
Results previous to the implementation of blanketing into the stellar atmospheres codes were discarded.
Targets without a determination of all [\Teff,\logg,\Rstar] parameters
were also removed from the cleaned sample.
Targets reported to have weak winds \citep{Pal96,Bal03,Mal04,Mal05b,Mal09,GNH11b}
are enclosed within a circle in Fig.~\ref{f:vinfZ} and are not included in the plots of the cleaned sample.

Known binary stars were also discarded.
However a non-negligible fraction of the clean sample 
could belong to undetected multiple systems.
We show the total sample in Fig.~\ref{f:vinfZ} to minimize the effect of this kind of bias.

The typical uncertainties of the parameters of the cleaned sample of LMC and SMC stars,
extracted from the references provided at the beginning of this section,
are: $\Delta$\Teff=1000~K, $\Delta$\logg=0.1dex, $\Delta$\Rstar=1\Rsun~ and $\Delta$\vinf=150\kms.
The ensuing error bars are provided within a box in Fig.~\ref{f:vinfZ}.
We note that these values underestimate the actual errors in two cases: 
(i) the terminal velocities of the latest O-dwarfs, with weaker winds, and
(ii) the effective temperature (hence gravity and \vesc) of the hottest O-stars (see Sect.~\ref{ss:vv}).
Because of error propagation, the actual error bars of \vinf/\vesc~ depend on the stellar parameters.
The vertical error bars of Fig.~\ref{f:vinfZ}-right have been calculated for two
stars with intermediate temperature:
 40000~K, \logg=3.6 and 25\Rsun~ resulting in $\Delta$(\vinf/\vesc)=0.50 for the supergiant plot, 
and 42000~K, \logg=4.0 and 10.5\Rsun~  resulting in $\Delta$(\vinf/\vesc)=0.34 for the dwarf star plot.
For IC~1613 stars we have used conservative uncertainties of
$\Delta$\Teff=2000~K and $\Delta$\logg=0.2dex \citep{GH13}.

\subsection{\vinf~ \textit{vs} Z}

Although the \Teff~ {\it vs} \vinf~ plot exhibits large scatter (Fig.~\ref{f:vinfZ}-left, rows 1 and 2),
it shows the known trend of increasing terminal velocity with increasing effective temperature \citep[e.g. ][]{KP00}.
The scatter is partly remedied when the cleaned sample is separated by luminosity class (Fig.~\ref{f:vinfZ}-left, rows 3 and 4)
as expected, since \vinf~ depends on \Teff~ through the escape velocity \citep{KP00}
which varies with \logg and the luminosity class.
The apparently decreased scatter in the B-supergiant regime is partly due to the slower terminal velocities 
of these objects.

An apparent flattening in the $\sim$33000-28000~K interval is observed in the total sample.
It disappears when only supergiant stars are considered, but then the \vinf-\Teff~ sequence 
seems discontinued at $\sim$33000~K.
The $\sim$33000-28000~K range marks the transition from O- to B-supergiants,
and the observed behavior could be the result of systematic
differences between the derived temperatures of both groups.
We note that \citet{VKL01} found an increase of mass loss rates at $\sim$35000~K
in the models of 1/30\Zsun, caused by the recombination of \ion{C}{4} to \ion{C}{3}, 
although the authors could not evaluate the exact impact on \vinf.
It is unlikely that this jump is related to the $\sim$33000-28000~K break
since the jump is only relevant in extremely metal-poor winds
where the main \Mdot~ drivers are CNO.

Regarding metallicity effects, the terminal velocities of Milky Way O-stars are
greater than the rest of Local Group counterparts
(weak wind stars being a clear exception to this statement).
The difference is more marked in the dwarf star sample, specially for \Teff$<$45000.
However, the location of IC~1613 O-stars
does not differ significantly from the position of LMC or SMC stars.
%
The only IC~1613 star with slower \vinf~ than the general trend (Fig.~\ref{f:vinfZ}-left, row 3) is \#63932,
the eclipsing binary.
%
%
Considering that MW dwarfs of \Teff=40000~K have \vinf=3000\kms~ on average
and using $\rm v_{\infty} \propto Z^{0.13}$,
we would expect that LMC/SMC/0.1 \Zsun~ stars had \vinf$\sim$2740/2430/2220\kms.
At least the expected difference between LMC and IC~1613 stars is larger than the quoted 
$\Delta$\vinf~ uncertainties.
This poses the question of whether there is an essential difference between the winds of
MW stars and those of $\leq$0.5\Zsun~ stars in the O-star regime.
%

We now evaluate separately the sample of dwarf and supergiant stars, 
to look for luminosity effects.
When all targets (all luminosity classes, all galaxies) are considered, there is an apparent saturation
effect at the highest temperatures \Teff$>$45000.
The effect disappears when the sample is broken between dwarfs and supergiants.
The dispersion in the plots  (Fig.~\ref{f:vinfZ}-left, rows 3 and 4) is still large, but the
terminal velocity of dwarf stars seems to depend more steeply on effective temperature
than the supergiant stars.
Note that if \#63932 had luminosity class~V, it would nicely follow the trend of the dwarf stars.

In the B-supergiant regime (\Teff$\lesssim$30000~K), there is a smaller number of studied targets.
No clear segregation of stars according to their host galaxy is observed.
\citet{Eal04c} arrived at similar conclusions in their comparison of
Galactic and SMC B-supergiants.

\subsection{\vinf/\vesc~ \textit{vs} Z}
\label{ss:vv}

The behavior of the \vinf/\vesc~ ratio as a function of temperature 
is shown in Fig.~\ref{f:vinfZ}-right.
The scatter is now more marked, mostly because of the additional 
uncertainties introduced by the escape velocity.
Until Gaia produces accurate distances, Galactic stars 
suffer from severe distance uncertainties which propagate to the
stellar radius and \vesc. 
Outside the Milky Way this is largely remedied, but then the problem is the
determination of gravities because the Balmer series is often hampered
by nebular contamination.
In addition the escape velocity and its
propagated error bars are very sensitive to the Eddington factor Gamma,
which grows rapidly for the most luminous and/or hottest stars
(see e.g. the errors of IC~1613 stars in Fig.~\ref{f:vinfZ}-right).

%

The location of MW stars in the right panels of  Fig.~\ref{f:vinfZ}
do not differ significantly from those of
lower metallicity counterparts.
When supergiant stars are considered separately
the \vinf/\vesc~ ratio follows the \citet{Cal06} and \citet{MP08} relation,
increasing with increasing temperature in the coolest types
and then flattening
around the canonical 2.65 value for \Teff$>$25000~K.
We note that our compiled data show a smooth transition from the low to high temperature regimes
of the bi-stability jump rather than an abrupt break.
The eclipsing binary \#63932 is again an outlier in this plot.

If only the dwarf stars are considered, the results are overall consistent 
-but do not clearly follow- the canonical value of 2.65 for the 
high temperature regime of the bi-stability jump. 
Instead, it would seem that the \vinf/\vesc~ ratio increases with increasing effective temperature,
although the scatter is too large to draw strong conclusions.

The five SMC dwarfs clearly under the main trend are 
      AV267  \citep[  O8V, 35700~K, \vinf/\vesc= 1.26, ][]{Bal13}, 
      AV440  \citep[  O8V, 37000~K, \vinf/\vesc= 1.15, ][]{PMal05},
      AV429  \citep[  O7V, 38300~K, \vinf/\vesc= 1.30, ][]{Bal13}, 
      AV446  \citep[O6.5V, 39700~K, \vinf/\vesc= 1.40, ][]{Bal13}, and
       AV14  \citep[  O5V, 44000~K, \vinf/\vesc= 1.66, ][]{PMal04}.
They have not been reported weak wind stars to our knowledge.
All have an intermediate \ion{C}{4} P~Cygni profile except AV440 and AV446
which show a blue-shifted wind trough.
Gravity (hence \vesc) may be unreliable for at least the three stars
analyzed by \citet{Bal13}, who could not constrain this parameter and adopted \logg=4.0.

The scatter of the hottest LMC dwarfs, which mark two different trends,
is remarkable.
The stars with large \vinf/\vesc, that favor a scenario of increasing \vinf/\vesc~ with \Teff~ are:
    R136-055   \citep[        O3V, 47500~K, \vinf/\vesc= 5.2, ][]{PMal04},       
    R136-033   \citep[        O3V, 47000~K, \vinf/\vesc= 5.3, ][]{PMal05},       
 LH81:W28-23   \citep[O3.5V((f+)), 47000~K, \vinf/\vesc= 4.4, ][]{Ral12b}, and   
  VFTS\#016    \citep[  O2III-If*, 50000~K, \vinf/\vesc= 5.9, ][]{Eal10}.        
VFTS\#016 \citep[30~Dor~016 in][]{Eal10} could be moved to the plot for supergiants, but it would depart from the general behavior of the sample even farther.
The targets compatible with the canonical \vinf/\vesc $\sim$ 2.65 are:
   Sk-67d211   \citep[  O3III(f*), 52500~K, \vinf/\vesc= 3.3, ][]{Wal04} , and  
       BI237   \citep[  O2V((f*)), 53200~K, \vinf/\vesc= 3.3, ][]{Ral12a}.      
All six stars exhibit well developed UV \ion{C}{4} profiles hence there was presumably no problem 
in deriving \vinf.

The O2-3 spectral types correspond to the hot temperature range where \ion{He}{1}$\lambda$4471 is weak or absent.
In this regime, \Teff~ cannot be confidently derived from the helium ionization balance,
and the ionization equilibrium of other metallic lines (usually nitrogen) should be used.
\Teff~ uncertainties propagate into \logg, which in turn repercute \vesc.
With this in mind,
the effective temperatures of R136-055 and R136-033, derived from FASTWIND analysis of HHe lines, are probably not reliable.
VFTS\#016 and Sk-67d211 were analyzed with CMFGEN using the ionization balance of nitrogen (optical lines)
and oxygen (ultraviolet lines); however, gravity was not derived but pre-assigned \logg=3.75 in the analysis,
and the subsequent escape velocities are not reliable. 
These examples illustrate the added uncertainties that
must be given careful consideration before interpreting the \vinf/\vesc~ \textit{vs} \Teff~ trends and scatter.

Effective temperatures and gravities for both BI237 and LH81:W28-23, on the other hand, were derived
using the latest implementation of the nitrogen model in the FASTWIND code by the same group of authors.
The two stars have different ratios: BI237: \vinf/\vesc=3.3, LH81:W28-23 \vinf/\vesc=4.4 .
Small number statistics prevent us from drawing any strong conclusions,
but it is clear that the 2.65 factor may lead to very different values for terminal
velocities than the actual one.

\subsection{Discussion in the framework of radiation driven wind theory}
Fig.~\ref{f:vinfZ} raises three main issues, which contradict the initial expectations
of this work:
\begin{itemize}
  \item[-]{IC~1613, SMC and LMC O-stars are not clearly separated in the \vinf~ or the \vinf/\vesc~ plots,
          and only depart significantly from MW stars in the \vinf~ \textit{vs} \Teff~ figure.}
  \item[-]{There are significant, non-systematic departures from the \vinf/\vesc$=2.65=constant$ relation,
widely used for \Teff$>$21000~K.}
  \item[-]{B-supergiants of all Local Group galaxies are found in the same locus.}
\end{itemize}
Are these results compatible with radiation driven wind theory?

Radiation line-driven wind theory for stationary, spherically
symmetric winds predicts that \vinf~ is roughly proportional to \vesc~
\citep[][Sect.4.1]{KP00}, following 
\begin{equation}
\label{eq:vesc}
v_{\infty} = 2.25 \, \alpha/(1-\alpha) v_{esc} \, f_1(\alpha) \, f_2(\delta) \,f_3(v_{esc})
\end{equation}
$\alpha$~ and $\delta$~ being the line force multipliers (LFM) and 
$f_1$, $f_2$~ and $f_3$~ functions of the order of unity.
%
Equation~\ref{eq:vesc} is derived from the simplified formulation of constant LFM through the wind.
The exponent of the power-law line-strength distribution, $\alpha$,
describes the behavior of the lines that drive the wind
and actually varies through the wind layers.
Its local value is subject to the species ionization balance,
and the abundance, temperature and density stratification. 
\citet{K02} allowed for varying LFM and showed that a relation similar to Eq.~\ref{eq:vesc} still holds, 
but the proportionality factor now depends on both metallicity and stellar luminosity.

Using Eq.~\ref{eq:vesc} or its depth-dependent-LFM analogue to estimate terminal velocities
would require detailed knowledge of the line acceleration of each star.
Instead, the empirical formula written by \citet{KP00} is most commonly used,
which prescribes \vinf/\vesc=2.65 for \Teff$>$21000~K.
The terminal velocity is then scaled by $\rm v_{\infty} \propto Z^{0.13}$~ \citep{LRD92} in non-solar metallicity environments.
Note that both Eq.~\ref{eq:vesc} and \citet{KP00}'s formula can explain the increasing trend of \vinf~ with \Teff;
the dependency is indirect, through the increasing trend of the escape velocity with effective temperature.

\citet{KP00}'s prescription is very useful in its simplicity but it is often forgotten that \vinf/\vesc=2.65 is a mean
value derived from a sample with large scatter, 
and that the authors estimated an accuracy of 20\% for the relation.
In addition, the used 
escape velocities were derived from
calibrations of stellar parameters with spectral type previous to the full implementation of line blanketing,
and from evolutionary masses.
\citet{Hal01} showed that when spectroscopic masses were used, the scatter decreased and the proportionality factor changed.
\citet{MP08} have recently reviewed this relation for Galactic supergiants using improved stellar parameters,
and obtained \vinf/\vesc=3.3 for \Teff$>$23000~K, but the reported dispersion is still large.
As a mean value for SMC stars both simulations \citep{K06} and observations \citep{Eal04c} have found that  \vinf/\vesc=2.63,
but both warn that a large scatter of \vinf/\vesc~ with \Teff~ is to be expected.

As we explain in the previous section, the observational scatter is partly caused by the
uncertainties involved in calculating escape velocities.
However, part of the scatter 
may be
real.
It follows from the nature of line acceleration that $\alpha$ (and the \vinf/\vesc~ proportionality factor)
may vary between stars of the same effective temperature if the luminosity, wind properties
and surface abundance patterns are different.
Small \Teff~ variations of 1000~K, within typical uncertainties,
may cause \vinf/\vesc~ variations as large as 17\%
\citep[see Fig.6 of][assuming \vesc=constant]{K06}.
This is comparable to our estimated error bars of $\Delta$(\vinf/\vesc)=0.50 (supergiants) and
$\Delta$(\vinf/\vesc)=0.34 (dwarfs),
which represent 18.9\% and 12.8\% of \vinf/\vesc=2.65.
Conversely, stars of different composition and luminosity may show similar \vinf/\vesc~
\citep[see Fig.~9 of ][]{K02}.
These points can explain both the observed scatter in Fig.~\ref{f:vinfZ} for stars
of the same host galaxy,
and why the stars from the LMC, SMC and IC~1613 are not clearly separated.

Even if the stellar element abundances and luminosity were well constrained,
there may be no easy parameterization for the \vinf/\vesc~ ratio.
\citet{K06} calculated the expected terminal velocities 
for a number of stars in the SMC, taking into account their element abundances when available.
The calculation was repeated after multiplying all element abundances by 1.5.
The resulting variations of the estimated terminal velocities (while \vesc~ is constant) were not always monotonic,
finding larger \vinf~ for some stars and smaller \vinf~ for others.
The differences could be up to $\sim$15\% for some targets, which translates into
\vinf/\vesc~ changes of the order of 10\%.
These variations do not fully account for the scatter seen in Fig.~\ref{f:vinfZ}-right,
but they again argue against using a single scaling factor to estimate terminal velocities from \vesc.

The explanation is that the metallicity variation altered the wind structure (T(r), $\rho$(r))
and ionization distribution (hence radiative acceleration)
and illustrates the delicate dependency of terminal
velocities on the conditions of the wind.
It is well known that the radiative force evolves through the wind \citep[e.g.][]{VKL99,PSL00,K06}.
Iron and iron-group elements dominate the inner layers,
and are therefore drivers of the mass loss rate.
The acceleration of the outer layers, which determines the terminal velocity,
is dominated by a small number of light elements that have strong resonance lines and
still keep their ionization stage farther out in the wind.
Because the outer wind is driven by a few dozen lines only, it is very sensitive to the local conditions. 
What \citet{K06}'s experiment showed is the high sensitivity of \vinf~
to apparently subtle differences between stars,
that can create very different conditions in the outer wind.
The poorly known metal content of stars 
(including elements like Ne or Ar, important for radiative acceleration),
and departures from Solar chemical mixtures,
may lead to miscalculation of the theoretical terminal velocities  and large departures
from the \vinf/\vesc~ average relation.


We conclude that the intrinsic errors of the stellar properties,
the small differences in stellar luminosities and abundances,
and the non-monotonic dependence of \vinf/\vesc~ on metallicity as a general parameter,
contribute significantly to the scatter seen in Fig.~\ref{f:vinfZ} and the lack of separation
of IC~1613, SMC and LMC O-stars.
The effect of additional factors that can affect \vesc,
such as the rotational velocity,
should also be explored.
An additional interesting question that should be explained is why MW stars are clearly 
separated in the \vinf~ \textit{vs} \Teff~ figures
(which could be interpreted as the Galactic stars departing
from the low metallicity, low density wind regime)
but do not clearly separate from the whole sample in the  \vinf/\vesc~ \textit{vs} \Teff~ charts.
This point can only be addressed with accurate Gaia distances 
and an effort to derive consistent parameters from a large homogeneous sample
such as the IACOB project \citep{SDal11a,SDal11b}.
\textit{We remark once again that \vinf/\vesc~ should be calculated on a star-by-star basis for the O-sample, losing its 
functionality as a quick means to estimate terminal velocities when accuracy is needed.}

Finally, the increasing \vinf/\vesc~ trend with increasing luminosity is mostly monotonic when the metallicity is constant.
\citet[Fig.~9 of ][]{K02} shows that at 0.2\Zsun~ metallicity, 
   \vinf/\vesc$\sim$2 for stars with $\log L/L_{\odot}$=6.30 but
   \vinf/\vesc$>$3    if             $\log L/L_{\odot}$=7.03.
Note that B-supergiant samples are usually biased towards the brightest objects of external galaxies.
The observed behavior of B-supergiants in Fig.~\ref{f:vinfZ} could be explained
by the \vinf/\vesc~ ratio (and \vinf) being dominated by the luminosity dependence.

\section{IC~1613's metal content}
\label{s:metal}

\begin{table*}
\caption{Metallicity measurements for IC1613}
\label{T:Z}
\centering
{\small
\begin{tabular}{l l l l}
\hline
\hline
 Target         &  12+log(O/H)        &  Z            &   $\rm [Fe/H]$          \\
\hline           
  \multicolumn{4}{l}{\it Stellar abundances from spectral analysis}\\ 
\hline
B-supergiants  &                         &               &                         \\
~~~~~ average  & 7.90 $\pm$ 0.08 (Bal07) &               &                         \\
~~~~~ min,max  & 7.80,8.00       (Bal07) &               &                         \\
M-supergiants  &                         &               & -0.67$\pm$0.09 dex (Tal07) \\
\hline
  \multicolumn{4}{l}{\it HII regions spectroscopy}\\ 
\hline
S3             & 7.86 $\pm$ 0.15 (T80)         &   & \\
               & 7.87 (DK82); 7.70 (KB95)      &   & \\   
               & 7.62, 7.88, 7.71 (LGH03)      &   & \\   
               & 7.72 $\pm$ 0.02 (Bal07)       &   & \\
               & 7.56 $\pm$ 0.11 (Tal13)       &   & \\
S8             & 7.83 (PBT88); 7.60 (DD83)     &   & \\   
A10 (S12)      & 8.84 (HG85)                   &   & \\ 
A17 (S10+S13)  & 8.87 (HG85); 7.80 (PBT88)     &   & \\
A13            & $>$ 7.61, 7.90, 7.89 (LGH03)  &   & \\   
S17            & 7.78 $\pm$ 0.05 (Bal07)       &   & \\
\hline
  \multicolumn{4}{l}{\it Evolutionary analysis: inferred initial abundance of WR}\\ 
\hline  
               &                     &   0.001 (KBS93)   &                          \\
               &                     &   0.004 (KB95)    &                          \\
\hline
  \multicolumn{4}{l}{\it Photometric studies, RGB tip}\\ 
\hline
               &                     &                   & -1.3 $\pm$0.8 dex (F88)  \\
               &                     &                   & -1.4 dex on average (Cal99) \\     
               &                     &                   & -1.75 dex         (TG02) \\
               &                     &                   & \textbf{-0.7 dex youngest, -1.3 dex oldest pop. (Sal03)} \\
               &                     &                   & \textbf{-1.3$\pm$0.1 dex on average (Sal14)} \\
\hline
\hline
\end{tabular}
\tablecomments{References- Bal07: \citet{Fal07}; Tal07: \citet{Tal07}; T80: \citet{T80};  DK82: \citet{DK82}; KB95: \citet{KB95}; 
Tal13: \citet{Tal13}; LGH03: \citet{LGH03}; PBT88: \citet{PBT88}; DD83: \citet{DD83}; HG85: \citet{HG85}; KBS93: \citet{KBS93}; 
F88: \citet{F88}; TG02: \citet{TG02}; Cal99: \citet{Cal99}; Sal03: \citet{Sal03}; Sal14: \citet{Sal14}.
}
}
\end{table*}



The low oxygen abundances measured in IC~1613
made this galaxy strategically important to study massive stars in the low metallicity 
regime, and a potential proxy for the first stars of the Universe.
We have compiled the determinations of IC~1613 metal content available in the literature 
in Table~\ref{T:Z}.
Let us focus on the values determined for the youngest
population of the galaxy: 
those from \ion{H}{2} regions,
B-supergiant stars and red supergiants (RSG).

A number of \ion{H}{2} regions have been analyzed in IC~1613,
including S8, the SN remnant,
and S3, the \ion{H}{2} region ionized by the galaxy's
oxygen Wolf Rayet.
O-abundances can be very sensitive to the method and calibration
used in the analysis, 
and this reflects on the scatter seen in 
Table~\ref{T:Z}\footnote{For a detailed discussion on the different methods
used to study IC~1613's \ion{H}{2} regions, we refer the
reader to \citet{Fal07} and \citet{Tal07}.}.
It was nicely illustrated by \citet{LGH03} who analyzed the same IC~1613 \ion{H}{2}
regions using three different methods and calibrations.
However, if \citet{HG85} results are not considered,
the abundances for IC~1613's \ion{H}{2} regions group around 
12+log(O/H)=7.80.
The high electron temperatures of IC~1613's \ion{H}{2} regions \citep{Fal07,Tal13}
further support a low oxygen abundance, as oxygen is the dominant
cooling agent of nebulae.
12+log(O/H)=7.80
is in fair agreement with \citet{Fal07}'s oxygen abundances
for IC~1613's B-supergiants: 12+log(O/H)=7.90$\pm$0.08 on average.
Scaled by the oxygen abundance of the sun
\citep[12+log(O/H)=8.69 ][]{Aal09},
these values correspond to 0.13\Zsun~ and 0.16\Zsun~ respectively.

However, in this paper we have collected evidence
suggesting that the iron content of IC~1613 is, at least,
similar to the SMC iron content.
Morphologically speaking, we showed in Section~\ref{s:mor} that 
IC~1613 stars resemble SMC analogue stars
with the same spectral type (i.e. presumably very similar stellar parameters),
but with more depleted Fe pseudo-continuum.
In fact, the UV spectra of IC~1613's O3-4V((f)) is more similar
to the LMC counterpart in this sense.
Quantitatively speaking,
the observed photospheric transitions of iron
between P~Cygni wind profiles
are not consistent with the TLUSTY models of 0.1\Zsun~ metallicity.
The observations rather resemble the models with 0.2\Zsun~ metallicity,
and in some cases those of 0.5\Zsun.
This is consistent with \citet{Tal07} results,
who determined that the iron abundance of IC~1613's RSG is [Fe/H]=-0.67
on average, or equivalently 0.21\Zsun.


A final determination of the iron content of IC~1613's 
blue massive stars must await quantitative analysis
of joint optical and ultraviolet spectra,
with careful consideration of each star's temperature,
gravity and microturbulence \citep[e.g. ][]{Mal04}.
If confirmed, however, this result partly solves
the conundrum of the strong winds found for O-stars in this galaxy \citep[][see intro]{Tal11,Hal12}.
If the iron content, main driver of mass loss, is not as poor as previously thought
the stars could drive a normal CAK wind.


Our findings indicate that the [$\alpha$/Fe] ratio may be sub-solar
in the youngest stellar population of IC~1613
([$\alpha$/Fe]=-0.10 using oxygen abundances from B-supergiants and 0.2Fe$_{\odot}$),
in agreement with \citet{Tal07}'s reported [$\alpha$/Fe]$\sim$-0.1
for the RSGs.
This apparently small figure is significant
once we consider that very different stellar populations span only 
a range of $\sim$1dex \citep[see][ Fig.~7]{Tal07}.
Local Group dwarf irregular galaxies show [$\alpha$/Fe] ratios 
that cluster around [$\alpha$/Fe]$_{\odot}$~ with some scatter,
with much lower values than MW stars
of similar iron content \citep[see review by ][]{TVT09}.
In particular, sub-solar values have also been found in 
Sextans~A \citep{KVal04}, WLM (\citet{Val03}, but also see \citet{Ual08}) and NGC3109 \citep{Hal14}.

Element abundance patterns and ratios are important clues to the 
chemical evolution of galaxies \citep[e.g. review by ][]{M04}.
The trend of the abundance ratios with galactic age
track the relative contributions of the different polluting agents
over time (e.g. SNe type Ia and II, massive stars winds,
low and intermediate mass pulsating AGB stars)
and depends heavily on the initial mass function
and on the duration and efficiency of the star formation episodes.
Oxygen is mostly produced by type-II supernovae (SNe), and is rapidly released
by the death of short-lived massive stars.
Iron is mainly produced by type-Ia SNe (secondarily by SNe-II), and is therefore more steadily
released into the ISM.
By computing the contribution of each population from present day abundances,
the  star formation history (SFH) of the galaxy is also  constrained.
However, this connection is not always biunivocal,
and the abundances can also be affected by gas flows (infall, outflow or internal flows).

In particular, IC~1613's low present-day [O/Fe] ratio 
indicates the absence of a dominant population of massive stars at recent ages,
which in turn can be explained by a number of scenarios.
It is compatible with star formation proceeding
in small and widely separated bursts \citep{GW91}:
in inter-burst periods the yields of SNe-II,
from massive stars which die fast,
stop being produced;
meanwhile, SN-Ia yields keep being produced and released.
More modern models of evolution with constant but low star formation rate can also
produce low [$\alpha$/Fe] ratios, if there are metal losses through galactic-scale winds \citep{TVT09}.
Alternatively, dwarf irregulars \citep[like dwarf spheroidals, ][]{TV05} may be unable to form very massive
molecular clouds and hence have a truncated initial mass function (IMF).
The detailed study of IC~1613's SFH by \citet{Sal03} found that star formation
was slow at early ages, perhaps suppressed by background radiation after the 
reionization of the Universe.
IC~1613 then experienced an extended period of comparatively enhanced (yet not intense) star formation rate
about 3 to 6 Gyr ago.
The authors updated work  \citet{Sal14}
finds a continuous, nearly constant SFH and again discards
any sharp increase of the star formation rate.
The fact that IC~1613 has not exhausted its gas argues in favor that 
it has not experienced any major or violent episode of star formation,
consistently with all the proposed scenarios.

\section{Summary and conclusions}
\label{s:fin}

This work presents the first HST-COS UV spectra of a sample of O- and early-B type stars in IC~1613.
It shows that UV spectroscopy of OB-stars beyond the SMC is feasible,
and that the COS spectra are apt for quantitative analysis. 
It opens new opportunities to characterize the winds of Local Group metal poor massive 
stars, needed as a proxy for the Population III.

Because they are technically challenging and expensive in observing time,
UV observations of OB-type stars outside the Milky Way are scarce.
However, UV spectroscopy is crucial to constrain
properly two key parameters for the winds of blue massive stars:
(i) it contains the best (and in most cases the only) direct diagnostics for the wind terminal velocity
and
(ii) it  provides direct observations of the iron content of O- and early-B type stars.

We obtained the terminal velocities of the target stars with the SEI method.
We found that at the S/N and spectral resolution of our dataset, it is important to use
realistic illuminating photospheric fluxes to properly account
for metallic lines that overlap with the bluest extent of the observed P~Cygnis.
We are confident on the derived terminal velocities within the estimated error bars, but
our analysis could not constrain well the turbulent velocity of the wind
and the $\beta$ exponent of the wind law.

The derived \vinf's constitute the first terminal velocity \textit{vs} spectral type scale
for a metal-poor galaxy beyond the Magellanic Clouds.
They also enabled the re-assessment of IC~1613's wind-momentum luminosity
relation.

We compared the terminal velocities of our IC~1613 targets with analogue stars
from the SMC, LMC and the Milky Way,
and analyzed the \vinf~ and \vinf/\vesc~ \textit{vs} \Teff~ dependence
using stellar parameters from modern works.
We found that IC~1613, SMC and LMC O-stars, although clearly separated from MW stars, 
do not occupy distinct locations in the \vinf~ \textit{vs} \Teff~ diagram.
In contrast, B-supergiants of all the evaluated Local Group galaxies overlap.
These results argue against a straightforward \vinf $\propto$ Z relation.
No segregation with host galaxy is seen in the  \vinf/\vesc~ \textit{vs} \Teff~ plot.

We also provide evidence that the iron content of our sample of IC~1613 stars may be 
higher than previously thought.
The abundance of iron and the metallicity had so far been scaled to $\sim$1/10 of the solar value
following the oxygen abundances of IC~1613's \ion{H}{2} regions 
\citep[which agree with the few available measurements from B-supergiants by ][]{Fal07}.
Note that this is a common practice since oxygen is one
of the most abundant elements in the Universe and its emission lines
can be easily detected and measured in weak nebulae and far away galaxies.
Results are pending confirmation, but the iron abundance of our targets seems closer
to the SMC value of 1/5 solar, in agreement with recent results from RSGs \citep{Tal07}.
In that case, the wind-momentum reported for IC~1613 stars by \citet{Tal11} would still be larger than expected from theory, 
but the discrepancy would decrease as there would be more metals to drive the wind.

Our results remark three well-known but often overlooked issues.
Firstly, oxygen should not be used blindly as a proxy for metallicity.
The abundance ratios of a galaxy, and in particular the [$\alpha$/Fe] ratio,
may differ significantly from the
solar value depending on its chemical evolution history.
[$\alpha$/Fe] is subsolar in IC~1613,
indicative of a quiet star formation history 
with no major episodes of star formation.

The exact mixture of elements, and in particular the abundance
of iron with respect to CNO, determines the stellar wind momentum 
and mass loss rate \citep{PSL00,VKL99,VKL01,K02,K06}.
Only in extremely metal poor stars (Z$\leq$1/30\Zsun), 
CNO elements drive the whole wind and mass loss \citep{VKL01}, and oxygen
can be safely used as an indicator for the metal content
when evaluating the WLR.
However, in more moderately metal-poor winds (Z$\leq$1/10\Zsun)
iron drives the inner flow and determines mass loss,
while the lighter elements drive the outer wind and determine 
the terminal velocity.
Their individual abundances must be separatedly known.

The second issue is that the terminal velocity 
cannot be safely estimated from the escape velocity.
We have found that Local Group early-type stars,
specially O-dwarfs,
significantly depart from the canonical \vinf/\vesc=2.65 relation \citep{KP00}
or the recently reviewed \vinf/\vesc=3.3 value \citep{MP08}.
Part of the scatter of the \vinf/\vesc~\textit{vs} \Teff~ diagram is due to
the uncertainties of the stellar parameters,
but part is real and expected in the framework of radiation driven wind theory.
As \citet{PSL00} explained, terminal velocities are extremely sensitive 
to apparently subtle effects, like small variations of temperature, density,
and chemical abundances,
that do alter the ionization balance of light elements in the outer wind. 
Any farther attempt to produce a simple empirical parameterization of \vinf/\vesc~
must use a large sample of stars for which a careful, homogeneous, tailored
determination of \Teff, \logg, \Rstar~ and stellar abundances is performed.

The final point to remark is that UV spectroscopy is 
absolutely necessary to characterize the winds of O- and early-B type stars.
As an example, our improved results allow us to re-evaluate the WLR for two targets,
\#62024 and \#69217.
Considering their likely SMC iron content,
their wind momentum calculated with the new terminal velocities 
is no longer in disagreement with theory.

%

\acknowledgments

Funded by Spanish MINECO under grants FIS2012-39162-C06-01, 20105Y1221, 
AYA2010-21697-C05-01, AYA2012-39364-C02-01, SEV 2011-0187-01 and 
by Gobierno de Canarias (PID2010119).
We would like to thank J. Puls, R.P. Kudritzki, and M. Cervi\~no for useful input.
This research has made use of the SIMBAD database,
operated at CDS, Strasbourg, France, 
the Aladin Sky Atlas \citep{Bal00},
and NASA's Astrophysics Data System.



{\it Facilities:} \facility{HST (COS)}



\appendix

\section{Sources of uncertainty of our SEI analysis}
\label{s:un}

The most direct source of uncertainty is the radial velocity correction,
which directly propagates into \vinf.
We have encountered wavelength calibration inconsistencies that render
the \vrad~ correction uncertain (see Sect.~\ref{ss:com}),
but they are mostly moderate ($\sim$ 50\kms) and
well enclosed within the error bars provided for \vinf.
The only exception is target \#62390, whose amplitude of discrepancy ($\sim$100\kms)
marginally exceeds the error bars of the derived \vinf~ ($\pm$75\kms).

Other potential uncertainty sources include spectral normalization and the use of an inappropriate
TLUSTY model as incident photospheric flux for the SEI method.
We elaborate on these issues in the following subsections.

\subsection{Normalization}
\label{ss:norm}

\begin{figure}
\centering
\includegraphics[height=\textheight]{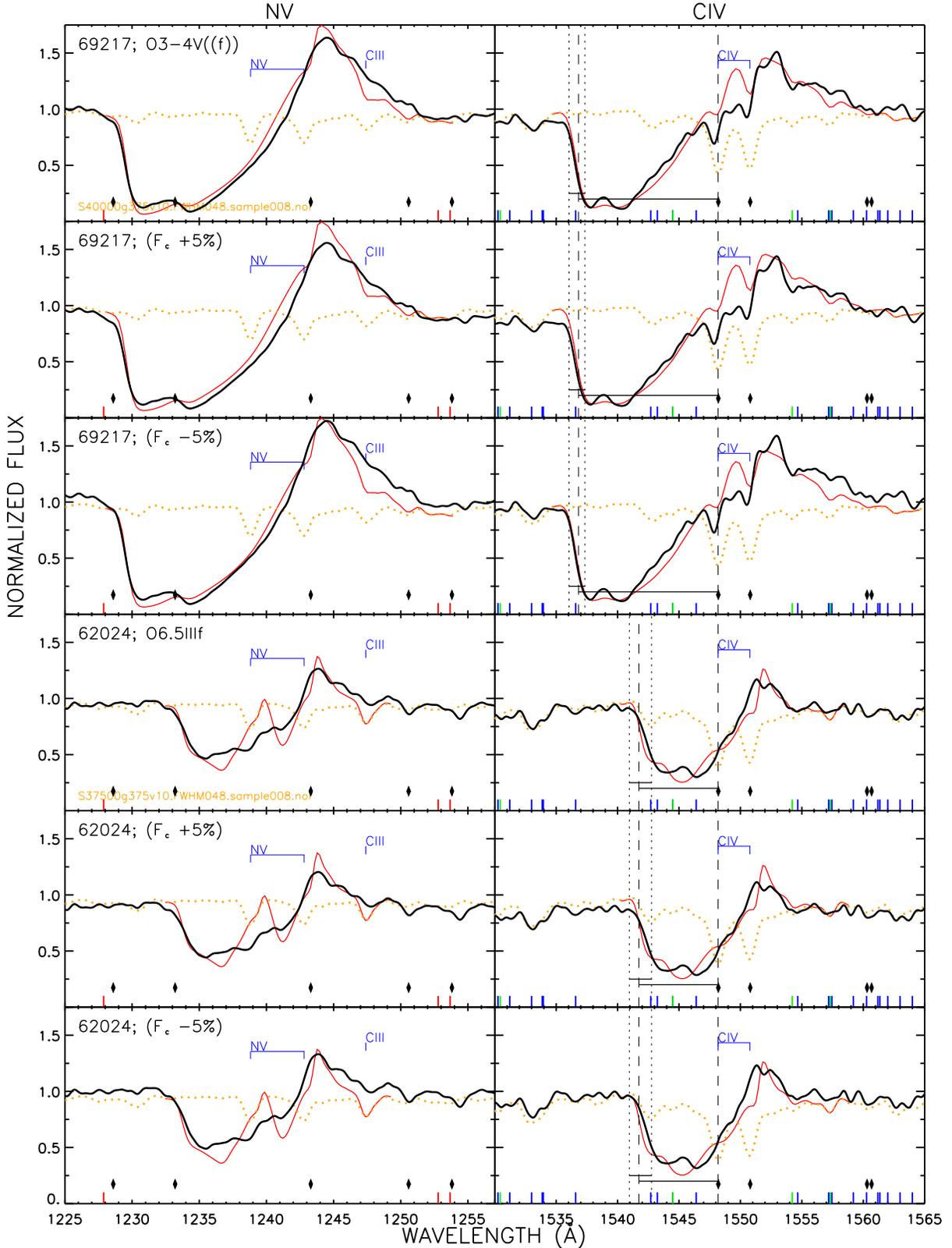}
   \vspace{-0.25cm}
   \caption{\footnotesize
Impact of the uncertainties introduced during the normalization
process on the derived terminal velocities for targets \#69217 and \#62024.
Colors and symbols as in Fig.~\ref{f:fitO}.
The terminal velocities derived for the targets are marked with vertical dashed lines,
and their errors with dotted lines.
For each star we show three plots. Firstly, its normalized spectrum at
the P~Cygni profiles of \ion{N}{5} and \ion{C}{4}, and the best fitting SEI profile. 
Then we normalized the stellar spectrum by a continuum curve raised
by 5\%, 
and show that it can be fit with the same SEI profile.
We repeat the process but decreasing the continuum curve
by 5\%.
           }
      \label{f:norm_vinf}
\end{figure}

\subsubsubsection{Quantifying the typical normalization error}


The spectra were normalized by fitting their continuum to a smooth
function so that the resulting normalized spectra matched a TLUSTY model.
This is a necessary simplification for the method to be practical, however, we have to quantify
how much the interplay of [\Teff, \logg, Z] varies the location of continuum points
and affects the normalization.

We evaluate these effects on \#69217, a target with good S/N.
Our preliminary joint optical+UV analysis with CMFGEN had determined \teff=42800~K, \logg=3.6 for this star,
but the TLUSTY model that best reproduces the Fe lines has  \teff=40000~K, \logg=3.75.
We contrasted a total of 18 normalized models from the TLUSTY grid with \\
  \teff=40000~K, \logg=3.5, 3.75 and 4.0, \\
  \teff=42500~K, \logg=3.75, 4.0 and 4.25, \\
and metallicity 0.1, 0.2 and 0.5\Zsun.
We searched for the wavelength intervals where the models nearly overlapped,
and defined continuum-boxes that enclosed all the synthetic spectra at the selected wavelengths.
Some boxes were discarded because of:
i) small problems in wavelength calibration,
ii) locally decreased S/N within the box,
iii) overlap with interstellar lines,
iv) overlap with P~Cygni profiles, and
v) overlap with Ly$_{\alpha}$~ wings.

The maximum height of the boxes
is a quantitative estimate of the error in our normalization.
We find that it amounts to 5\% of the total continuum at most.  
We therefore adopt $\pm$2.5\% as the typical error of our normalization.

As a consistency check, we fitted the continuum boxes to a 3rd-degree polynomial
(excluding Ly$_{\alpha}$ and shorter wavelengths with decreased detector sensitivity)
and generated a new normalization function.
It meets the normalization function used in Sect.~\ref{ss:nor} within the listed 5\%.
%

\subsubsubsection{Impact on \vinf}

Fig.~\ref{f:norm_vinf} illustrates that the uncertain normalization
does not affect the terminal velocities derived in this work.
We rectified \#69217 by two continuum curves obtained by raising and decreasing
our initially determined normalization curve by 5\%
(twice the error estimated in the previous section, to be on the safe side).
The best fitting SEI profiles for \#69217's original \ion{C}{4} and \ion{N}{5} P~Cygnis
match within the error bars the resulting profiles with the new normalization.

We repeated the test on a star with clearly unsaturated P~Cygni profiles: \#62024.
Incidentally, the amplified plot of Fig.~\ref{f:norm_vinf} better illustrates how 
the overlapping \ion{Fe}{4} $\sim$1542.7\AA~ line affects the \ion{C}{4} fitting.  
Again, the SEI profile that best reproduces the original data
agrees within the error bars with the re-normalizations.
The small detected differences can also be absorbed by the turbulent velocity $v_{to}$,
which is one of the unconstrained parameters of this work.
We conclude that considering the resolution and S/N of the observations and
the overlap of metallic lines, 
the effect of the normalization on the derived terminal velocities
is negligible.

\begin{figure}
\centering
\includegraphics[width=\textwidth]{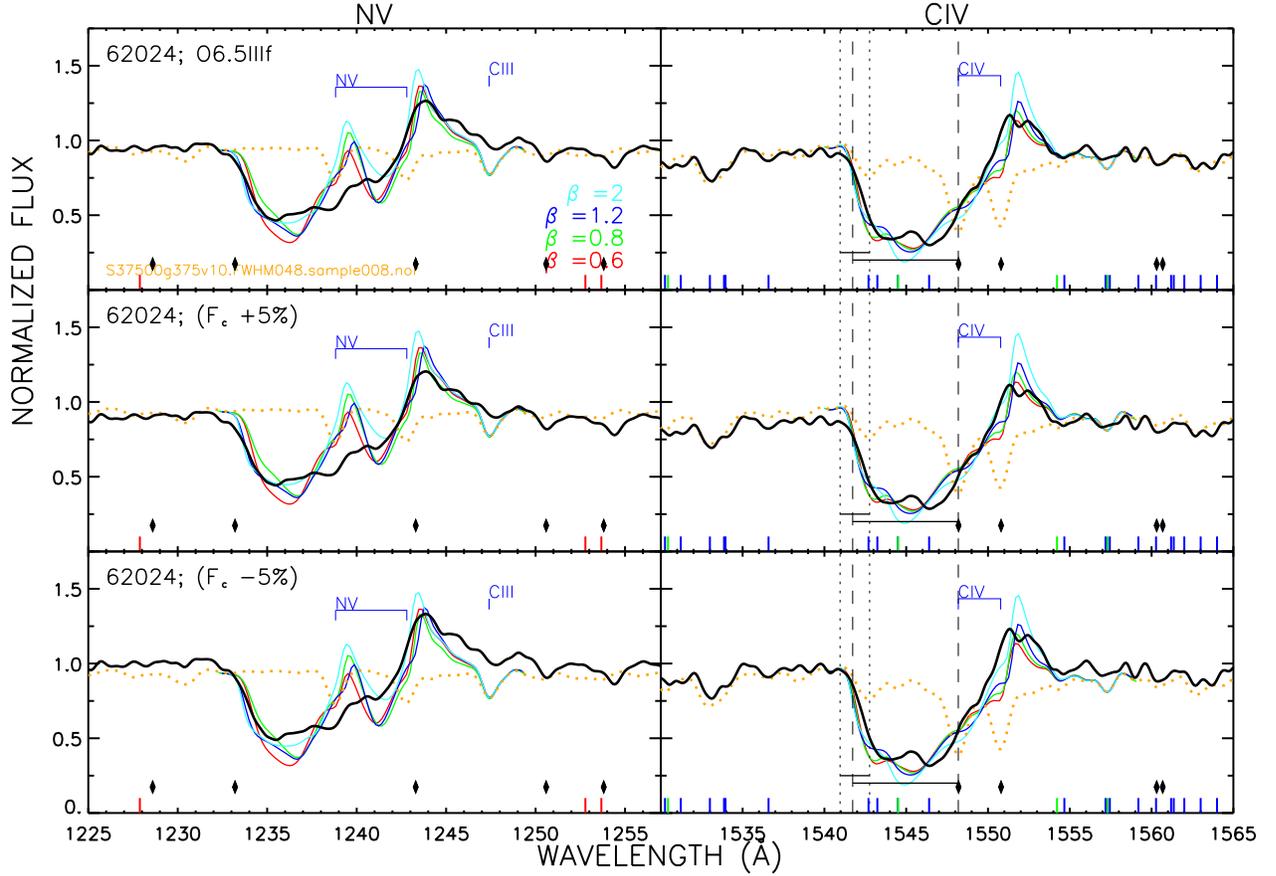}
   \caption{Sensitivity of our analysis to the $\beta$~ parameter.
The structure of the plot is the same as in Fig.~\ref{f:norm_vinf}.
Colors represent SEI-generated profiles with different exponents
of the velocity law, as indicated in the plot. The terminal and turbulent
velocities were also varied, to fit the observed profile.
The top panel shows that
an equally good fit can be achieved for any $\beta$=0.6-1.2,
considering the spectral quality.
For this star we chose $\beta$=1.2 as the best fit to \ion{C}{4}.
In the second/third panel, the continuum curve for normalization
was increased/decreased by 5\%.
The normalization by the increased curve favors smaller values of $\beta$~
and \textit{vice versa}, although the quality of the fit does not significantly improve
for any of them.
           }
      \label{f:norm_beta}
\end{figure}

\subsubsubsection{Impact on $\beta$}

As we explained in Sect.~\ref{s:sei},
the combination of several factors prevents our analysis 
to provide firm constraints on $\beta$.
We evaluate the impact of a poor normalization in
Fig.~\ref{f:norm_beta}.
If the normalization yields a lowered continuum, the P~Cygni emission is also smaller,
and the analysis will determine smaller $\beta$'s. On the other hand, if the stellar continuum
is raised, a larger $\beta$ will be derived.
Note that the quality of the fit only changes marginally.

\subsubsubsection{Impact on Z}

\begin{figure}
\centering
\includegraphics[height=0.5\textheight,angle=90]{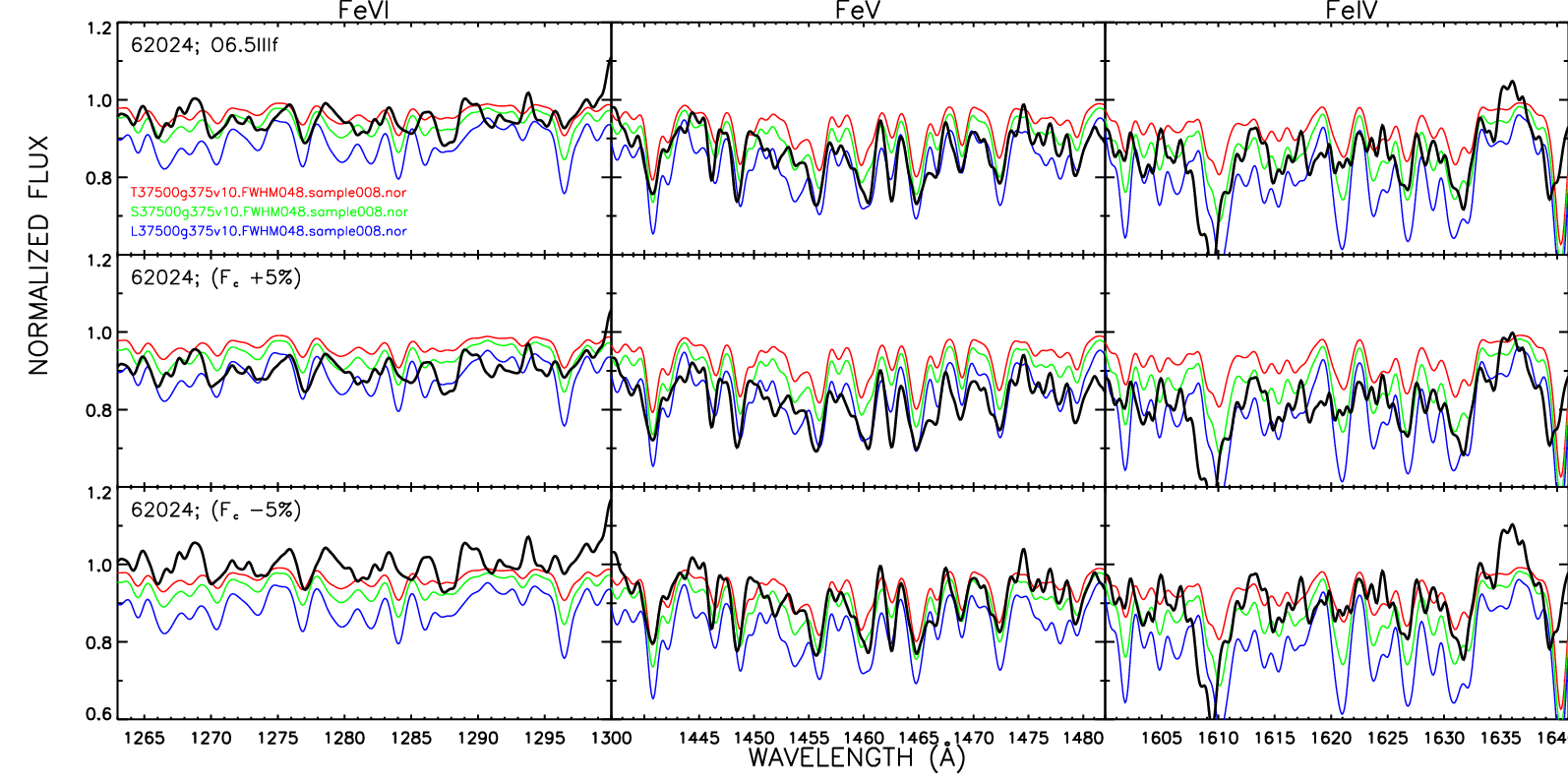}
   \caption{Impact of normalization
on the estimated metallicity of target \#62024.
The structure of the Figure is the same as Fig.~\ref{f:norm_vinf},
but now showing pieces of the stellar continuum.
TLUSTY synthetic spectra with the same \teff~ and \logg~
as the reference TLUSTY model but varying metallicity are shown
(Red: 0.1\Zsun, green: 0.2\Zsun, blue: 0.5\Zsun).
           }
      \label{f:norm_metal}
\end{figure}

Contrastingly, normalization has an important impact on the derived metallicities, as reflected in
Fig.~\ref{f:norm_metal}.
The plot compares \#62024's spectrum with
TLUSTY spectra with the same \teff~ and \logg
as the model used for analysis (Table~\ref{T:vinf})
but varying metallicity (0.1, 0.2 and 0.5\Zsun).
The synthetic spectra are compared to the three normalizations explained above.
Fig.~\ref{f:norm_metal} shows the intervals corresponding to the transitions of \ion{Fe}{4}, \ion{Fe}{5} and \ion{Fe}{6}
to ensure that there are no temperature effects
by examining the three ionization stages simultaneously.

The model that overall best reproduces the Fe lines of \#62024 
has SMC metallicity.
This is shown in Fig.~\ref{f:norm_metal}'s close-up, where the SMC model 
fits best the spectral features in the ranges dominated by \ion{Fe}{4}, \ion{Fe}{5} and \ion{Fe}{6}.
Fig.~\ref{f:norm_metal} also illustrates small mismatches between some (but not all) model and observed features.
The disagreement can be caused by the reported small problems with the wavelength calibration or
by inaccuracies in TLUSTY's Fe atomic model.

When the absolute spectrum is normalized by a continuum curve raised by 5\%
and compared to TLUSTY models,
the \ion{Fe}{6} and \ion{Fe}{5} ranges would rather suggest that the metallicity is closer to the LMC.
Conversely, when the curve lowered by 5\% is used for normalization,
the metallicity inferred from \ion{Fe}{5} would be intermediate between SMC and 1/10 \Zsun.
However, note that we would likely discard this normalization as a bad one.

In conclusion, UV metallicity determinations are very sensitive to the normalization process.
The iron content of resolved stars can be better derived from the analysis of the absolute flux,
in a joint process with the rest of stellar parameters as we explained in Sect.~\ref{ss:tlusty}.
We also advice caution on the use of spectral indexes to infer
the metallicity of unresolved populations from normalized spectra \citep[see e.g. discussion in ][]{Ral04}.

\subsection{Use of a reference TLUSTY model with inappropriate parameters}


The TLUSTY model assigned to carry out the SEI analysis is the
one that best reproduces the stellar pseudo-continuum.
However, because accurately constraining the stellar parameters requires a
much more detailed analysis
and anyway we are limited by the parameter space covered by the grid,
the model does not always offer a good representation.
We evaluate in this section how selecting the wrong TLUSTY model
can affect the derived wind parameters.

Either we err on the stellar parameters [\Teff,\logg,$\xi$]
of the model or its metallicity,
the problematic outcome is that the strength of the observed Fe features is
not well reproduced.

The pseudo-continuum of target \#62024 was represented in the SEI analysis
by the TLUSTY model with SMC metallicity, \Teff=37500~K and \logg=3.75.
We have calculated SEI profiles with the same wind parameters
but two extreme models where \ion{Fe}{5} is produced 
clearly in excess (35000~K, \logg=4.0, 0.5\Zsun)
and almost absent (40000~K, \logg=3.50, 0.1\Zsun)
as input photospheric flux.

\begin{figure}
\centering
\includegraphics[width=\textwidth]{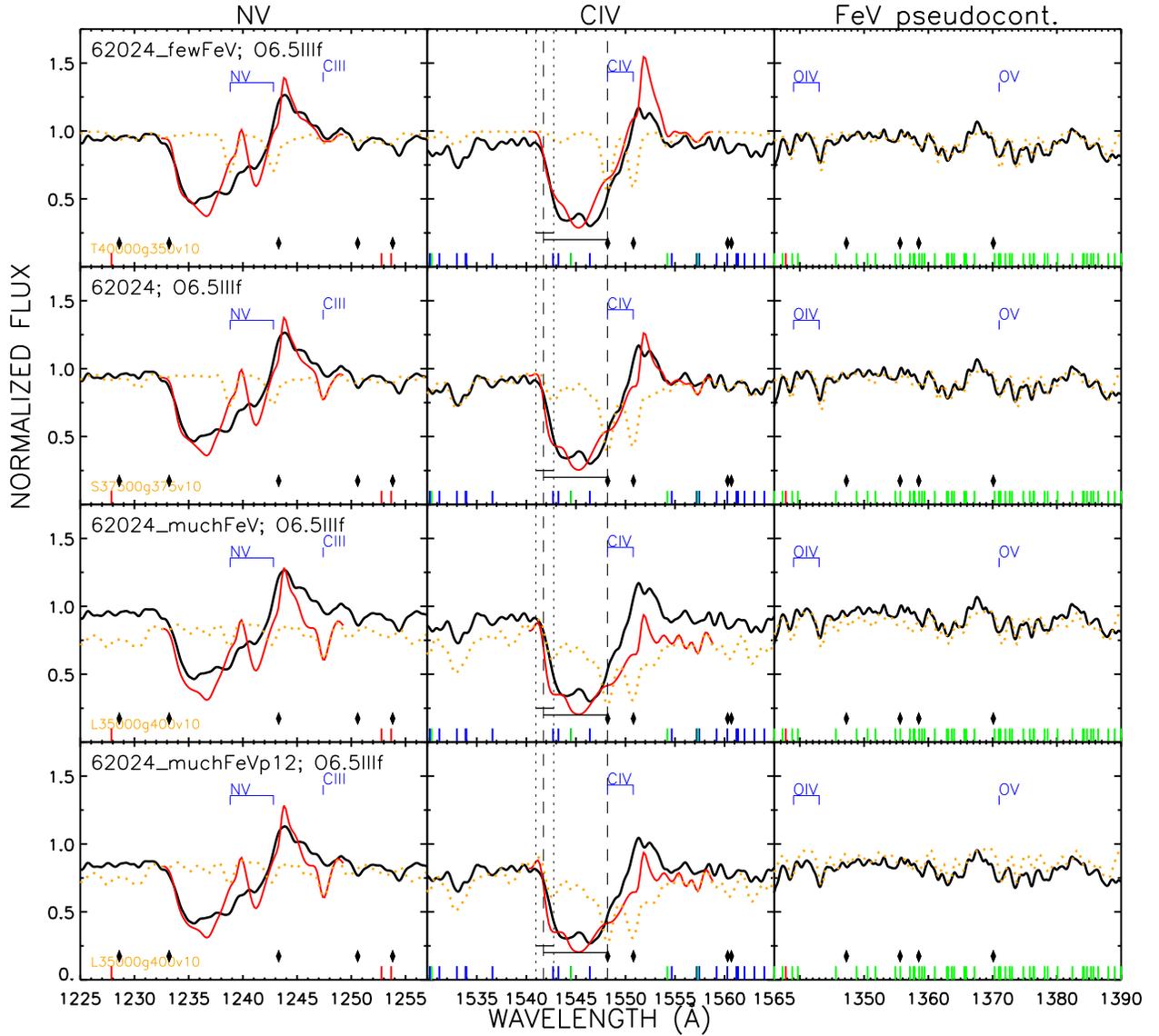}
   \caption{Impact of using an inappropriate TLUSTY model on
the resulting SEI profile and on the derived wind parameters.
Colors and symbols as in Fig.~\ref{f:fitO}.
The panel labeled 62024 shows the spectrum of \#62024
with the best fitting SEI profile and the selected TLUSTY model.
We show the same observations on top, but compared with a TLUSTY
model with very poor \ion{Fe}{5} content and the profile generated with the SEI code
using this model as input photospheric flux.
A similar exercise is performed underneath,
now with a model with \ion{Fe}{5} in excess.
In sight of the latter, we would probably re-normalize the spectrum
to lower the continuum; this is shown in the bottom panel.
We note that we would derive the same terminal velocities (within error bars) for the four cases.
           }
      \label{f:wrong}
\end{figure}

The resulting SEI profiles are shown in Fig.~\ref{f:wrong}.
Note that if the metal-rich model was used as reference,
we would have probably lowered the stellar continuum.
This is also shown in Fig.~\ref{f:wrong}.

We will focus our discussion on the \ion{C}{4}$\lambda\lambda$1548.2,1550.8 doublet,
affected by the strength of the \ion{Fe}{5} lines.
The profile calculated with the metal-poor model produces the P~Cygni emission 
in excess.
When the metal-rich model is used, the SEI code does not produce enough emission,
even after re-normalizing the spectra.
Note that if less extreme models were used, the discrepancy between the
observations and the profiles would not be as marked and could be solved
with a different value for $\beta$.

Contrastingly, in both cases we would derive the same terminal velocities (within
error bars) as if using the correct TLUSTY model.
A small correction for the turbulent velocity $v_{to}$~
would be needed for the metal-rich model, likely caused by the 
comparatively stronger \ion{Fe}{4}$\sim\lambda$1542.7\AA~ line.

This experiment highlights the strength and weakness of our analysis.
$\beta$~ is very sensitive to the normalization process, and to the 
reference TLUSTY model.
The metallicity estimate also depends on normalization,
but it is reliable enough in the frame of the analysis presented in Sections \ref{ss:mm} and \ref{ss:tlusty}.
The SEI method, on the other hand, is quite robust to determine terminal velocities.

%






\end{document}